\makeatletter 

\documentclass[aps,twocolumn,pra,tightenlines,floatfix,showpacs,superscriptaddress]{revtex4-1}

\usepackage{graphicx}
\usepackage{epstopdf}
\usepackage[english]{babel}
\usepackage{amsmath}
\usepackage{mathtools}
\usepackage{ntheorem}
\theoremseparator{:}
\theoremheaderfont {\it}

\usepackage{amssymb}
\usepackage{times}
\usepackage{appendix}
\usepackage{bm}
\usepackage{mathrsfs}   
\usepackage{float}
\usepackage{color}
\usepackage{longtable}
\usepackage{url}
\usepackage[usenames,dvipsnames]{xcolor}
\usepackage{subfigure}
\usepackage[percent]{overpic}
\renewcommand{\@thesubfigure}{\hskip\subfiglabelskip}
\usepackage{indentfirst} 
\graphicspath{{./Figures}}

\newcommand{\Rmnum}[1]{\expandafter\@slowromancap\romannumeral #1@}

\usepackage{array}
\usepackage{booktabs}

\newcommand{\rmi}{\text{i}}
\newcommand{\rme}{\text{e}}

\makeatother 

\usepackage[colorlinks=true,linkcolor=Blue,urlcolor=Blue,citecolor=Blue,anchorcolor=Red]{hyperref}
\usepackage{cleveref}
\crefname{equation}{Eq.~}{Eq.~}
\crefname{figure}{Fig.~}{Fig.~}

\begin{document}

\title{Dynamical multipartite entanglement in a generalized Tavis-Cummings model with $XY$ spin interaction}

%

%

\author{Yuguo Su}


\author{Zhijie Sun}

\author{Yiying Yan}

\author{Hengyan Wang}

\author{Junyan Luo}

\affiliation{School of Science, Zhejiang University of Science and Technology, Hangzhou 310023, China}

\author{Tiantian Ying}

\email{yingtt@zjut.edu.cn}

\affiliation{College of Civil Engineering, Zhejiang University of Technology, Hangzhou 310014, China}

\author{Hongbin Liang}
\email{lhb@cuz.edu.cn}
\affiliation{College of Media Engineering, Communication University of Zhejiang, Hangzhou 310018, China}

\author{Yi-Xiao Huang}
\email{yxhuang@zust.edu.cn}
\affiliation{School of Science, Zhejiang University of Science and Technology, Hangzhou 310023, China}

\begin{abstract}
Multipartite entanglement is a long-term pursuit in the resource theory, offering a potential resource for quantum metrology.
Here, we present the dynamical multipartite entanglement, which is in terms of the quantum Fisher information, of a generalized Tavis-Cummings (TC) model introducing the $XY$ spin interaction.
Since our model cannot be solved exactly, we theoretically derive and numerically examine the effective description of our model.
By the Holstein-Primakoff transformation, we show the bridge from the generalized TC model to the central spin model.
Furthermore, the reduced density matrix of the central spins is presented, which is the prerequisite for calculating multipartite entanglement.
We also discuss the effect of the temperature, the coupling constant, and the magnetic field on the dynamical multipartite entanglement in the central spin model, where the central spin is initially unentangled.
Strong coupling and low temperature are necessary conditions for a genuine multipartite entanglement in the $XY$ model, and together with the magnetic field, they govern the modulation of both the entanglement period and amplitude.
Our results unveil the deep link between the TC model and the central spin model, allowing for a better comprehension of their dynamical multipartite entanglement.

\end{abstract}
\maketitle


\section{Introduction}\label{Sec.I}
Nonlocal correlations and entanglement between distinct or separated physical objects are the most counterintuitive of quantum mechanics~\cite{PhysicsPhysiqueFizika.1.195,RevModPhys.86.419,RevModPhys.81.865}.
Entanglement serves as the key resource of quantum information~\cite{RevModPhys.90.035005}, whose practical usefulness is exposed as the cornerstone of quantum technologies, encompassing areas such as quantum sensing~\cite{RevModPhys.89.035002,PhysRevA.97.032329,Carollo_2019,10.21468/SciPostPhys.13.4.077,PhysRevLett.131.150802}, quantum computing~\cite{PhysRevLett.82.1784,PhysRevLett.88.097904,PhysRevA.85.062318}, quantum cryptography~\cite{PhysRevLett.67.661}, and quantum communication~\cite{PhysRevLett.69.2881,PhysRevLett.70.1895,PhysRevLett.76.722}.
To reach the long-term pursuit to achieve scalable multipartite entanglement, significant advances have been made in manifold quantum systems encompassing superconducting circuits~\cite{PhysRevLett.122.110501,doi:10.1126/science.aay0600}, atomic qubits in optical lattices~\cite{PhysRevLett.120.243201,science.aaz6801}, trapped ions~\cite{PhysRevLett.106.130506,PhysRevX.8.021012}, and neutral atoms in tweezer arrays~\cite{Omran2019,Graham2022}.
Therefore, endeavors to identify and quantify multipartite entanglement have the potential to reveal new phenomena and enhance our capability to harness their merits.
Entanglement witness is a reliable strategy that depends on determining the expectation values of operators and its selection relies on the type of system and entanglement of interest.
However, a prominent quantity is the quantum Fisher information (QFI)~\cite{PhysRevD.23.1693,PhysRevLett.72.3439,MA201189,Liu_2019,PhysRevB.108.144414,WOS:001170772600001}, which reflects the highest precision that can be achieved for an unknown parameter in quantum metrology.

Cavity quantum electrodynamics (Cavity-QED)  stimulates intense theoretical and experimental investigations in the study of a variety of physics, ranging from the light-matter interaction~\cite{Dutra-Book:2005,RevModPhys.73.565} to quantum entanglement~\cite{doi:10.1063/1.882326,RevModPhys.73.565}, coherence dynamics~\cite{doi:10.1126/science.1078446,WOS:000223746000038}, and other quantum-classical phenomena~\cite{PhysRevLett.76.1800,mckeever_experimental_2003}.
Cavity-QED provides an ideal platform to simulate cavity-atom interaction from weak to strong that makes for quantum computing~\cite{Bennett2000,Knill2001,Buluta_2011}, quantum simulation~\cite{Buluta2009,Georgescu2014}, quantum metrology~\cite{Giovannetti2011,YuguoSu2021,PhysRevA.109.042614,PhysRevLett.130.170801,deng2023,PhysRevLett.133.110201,PRXQuantum.5.030338,su2024approaching}, and quantum key distribution~\cite{Scarani2009,obrien_photonic_2009}.
In particular, the cavity-QED also produces promising applications in precision measurements~\cite{WOS:000368673800032,PhysRevLett.116.093602,Flower_2019,Gietka_2021} and quantum-enhanced sensing~\cite{Kim1999,WOS:000263818900002,PhysRevA.94.022313}.
With the initial impetus of the development provided by cavity-QED, circuit QED has since emerged as a distinct and flourishing field of research, including light-matter interaction~\cite{niemczyk_circuit_2010,PhysRevX.2.021007,RevModPhys.93.025005,lledo_cloaking_2023}, quantum information processing technology~\cite{clarke_superconducting_2008,Wendin_2017,10.1063/1.5089550,blais_quantum_2020}, and novel hybrid quantum systems~\cite{RevModPhys.85.623,clerk_hybrid_2020}.
The Tavis-Cummings (TC) model~\cite{PhysRev.170.379,PhysRev.188.692}, where nonrotating wave terms are absent, is known as the generalized model of $N$ two-level atoms occupying the same site and collectively interacting with the quantized single mode of the field.
To exploit the many-body effects, we propose a generalized TC model where the atoms are organized into a spin chain with the $XY$ interaction and map it onto a central spin model~\cite{gaudin_m_diagonalisation_1976,NVProkof'ev_2000,PhysRevLett.91.246802,RevModPhys.76.643,PhysRevLett.120.090401,PhysRevA.102.052423}.

In this work, the dynamical multipartite entanglement of a generalized TC model that involves the $XY$ spin interaction, is observed in terms of the QFI.
We theoretically derive and numerically examine the effective description of our model to avoid the difficulty of exactly solving it.
The generalized TC model can be mapped onto the central spin model by the Holstein-Primakoff (HP) transformation.
Moreover, we present the reduced density matrix of the central spins, which is the prerequisite for calculating multipartite entanglement.
We also analyze the effect of the inverse temperature $\beta$, the coupling constant $\eta$, and the magnetic field $h$ on the dynamical multipartite entanglement with an initially unentangled central spin.
Strong coupling is a necessary condition for the existence of entanglement, but it must be accompanied by low temperatures to achieve a genuine multipartite entanglement.
Additionally, they and the magnetic field dominate the modulation of the entanglement period.
Our results contribute to deepening your understanding of dynamical multipartite entanglement and pave the way for future applications in quantum metrology.


\begin{figure}[htbp]
	\centering
	\begin{minipage}{1\linewidth}
		\centering
		\begin{overpic}[width=\linewidth]{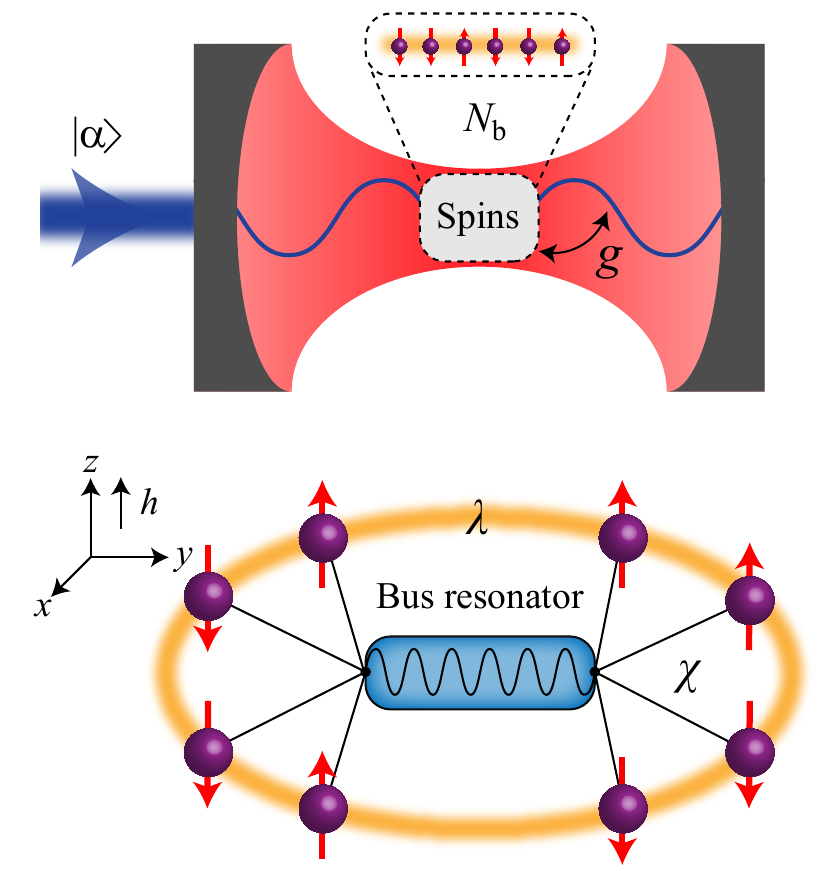}
			\put(0,92.5){(a)}
			\put(0,47.5){(b)}
		\end{overpic}
\end{minipage}
\caption{(Color online) 
	(a) Cavity-QED setup.
	A coherent field $\left|\alpha\right\rangle$ is injected into the cavity and is coupled to $N_\text{b}$ trapped spins (purple) with the coupling strength $g$.
	(b) Schematic illustration of the $N_\text{b}$-qubit superconducting quantum processor.
	The qubits are arranged along a circular chain via the nearest-neighbor couplings (orange),  denoted as the interaction strength $\lambda$.
	The bus resonator, which acts as a cavity field, couples equally to each individual qubit as illustrated by black lines (interaction strength $\chi$). Also see~\cite{doi:10.1126/science.aay0600,PhysRevLett.120.050507}.
}\label{Fig1}
\end{figure}

This paper is organized as follows.
In Sec.~\ref{Sec.II}, we concretely introduce the cavity-QED system and the HP transformation.
We derive the effective Hamiltonian obtained by the time-averaged method and demonstrate its validity by comparing the results of the original and effective one.
In Sec.~\ref{Sec.III}, we show the reduced density matrix of the central spin system, which could describe the time evolution of numerous intriguing quantities including multipartite entanglement.
We investigate the dynamical multipartite entanglement of the central spin system for the $XX$ model and the Ising model in Sec.~\ref{Sec.IV}.
We the impacts of the inverse temperature $\beta$, the coupling constant $\eta$, and the magnetic field $h$ on it, respectively.
Finally, a summary is given in Sec.~\ref{Sec.V}.

\section{System and HP transformation}\label{Sec.II}
We consider an atom-light coupling system describing the coupling of a single bosonic cavity mode to a collection of $N_\text{b}$ two-level atoms (or spin-1/2 spins),  as depicted in \cref{Fig1}:
\begin{eqnarray}
&&H=\omega_0 J_z+\omega_{\text{a}} a^\dagger a+H_0\left(h_{0}\right)+H_{\rm I},\label{H-1}\\
&&H_{0}\left(\!h_{0}\!\right)=\!-\frac{1}{2}\!\sum_{i=1}^{N_\text{b}}\!{\!\left\{\!\frac{\lambda}{2}\!\left[\left(\!1\!+\!\gamma\right)\sigma^x_i\sigma^x_{i+1}\!+\!\left(\!1\!-\!\gamma\right)\sigma^y_i\sigma^y_{i+1}\!\right]\!+\!h_{0}\sigma^z_i\!\right\}},\nonumber\\
&&H_{\rm I}=g\left(a^\dagger J_-+aJ_+\right).\nonumber
\end{eqnarray}
Here, $H_{\rm I}$ denotes the interaction Hamiltonian and $J_{x,y,z}=\sum_{j=1}^{N_\text{b}}{\sigma_j^{x,y,z}/2}$, $J_\pm=J_x\pm \rmi  J_y$ are the collective spin operators; $a^\dagger$ ($a$) is creation (annihilation) operator for the cavity mode; $\omega_0$, $\omega_{\text{a}}$ and $2g$ are the spin transition frequency, the cavity frequency and the single-photon Rabi frequency, respectively.
The $XY$ Hamiltonian $H_0\left(h_{0}\right)$ describes the interaction between the spins, where $\lambda$ is the nearest neighbor interaction, $\sigma^{\alpha}_{i}$ denotes the Pauli matrix ($\alpha=x,y,z$) on site $i$, $N_\text{b}$ is the number of sites, $\gamma$ the degree of anisotropy, and $h$ a transverse field.

Utilizing the unitary transformation $U=\exp\left\{-\rmi \left[\omega_0 J_z+\omega_{\text{a}} a^\dagger a+H_0\left(h_{0}\right)\right]t\right\}$ 
and the high-frequency approximation for spins ($\left|\omega_{\text{a}}+h_{0}-\omega_0\right|\gg\lambda$), the original Hamiltonian (\ref{H-1}) in the interaction picture could be read as 
$H_{\text{int}}= g\left[J_-a^\dagger\rme^{-\rmi \left(\Delta+\delta\right) t}+\text{H.c.}\right]$ with a large effective detuning $\Delta=\omega_0-h_{0}-\omega_{\text{a}}$ and a small field-irrelative residue $\delta$.
Employing the time-averaged method in the Ref.~\cite{James2007}, the total Hamiltonian can be approximated as (see more details in Appendix~A):
\begin{equation}\label{H2}
H\simeq H_0\left(h_{0}-\omega_0\right)+\frac{2g^2}{\Delta}J_za^\dagger a+\frac{g^2}{\Delta}J_+J_-,
\end{equation}
which is written in the Schr\"{o}dinger picture and the photon number is conserved.
The time-averaged method could be regarded as a natural generalization of the rotating-wave approximation~\cite{TAM} since its core is to obviate high-frequency contributions.
Alternatively, the Fr{\"o}hlich-Nakajima transformation is applied to derive the effective Hamiltonian~\cite{MA201189}, with the validity of the approximation in Eq.~\eqref{H2} ensured by the large-detuning condition, i.e., $\Delta\gg g\sqrt{N_{\text{b}}}$.

Experimentally, the average photon number is much larger than the number of up spins, i.e., $\langle a^\dag a\rangle\gg \left\langle J_z\right\rangle= \left\langle \left[J_+,J_-\right]\right\rangle/2\approx N_{\text{b}}$, then we can further reduce the Hamiltonian (\ref{H2}) to obtain the final effective Hamiltonian: 
\begin{equation}
H_{\text{eff}}=H_0\left(h_{0}-\omega_0\right)+\frac{2g^2}{\Delta}J_za^\dagger a.\label{H_eff}
\end{equation}
Under the same assumption $\langle a^\dag a\rangle\gg 1$, by using the HP transformation $S_+=\sqrt{N_\text{c}}a^\dagger\sqrt{1-a^\dagger a/N_\text{c}}$, $S_-=\sqrt{N_\text{c}}\sqrt{1-a^\dagger a/N_\text{c}}a$ and $S_z=-N_\text{c}/2+a^\dagger a$, the cavity-QED system can be mapped onto a central spin system where cavity photons correspond to $N_\text{c}$ central spins.
The corresponding Hamiltonian calculated from (\ref{H_eff}) reads
\begin{equation}
H_{\rm CSM}=H_0\left(h\right)-2\eta J_z S_z
,\label{CentralSpinContext}
\end{equation}
where $\eta=-g^2/\Delta^2$ is a coupling constant, $S_{z}$ can be treated as the collective spin operator of the central spins, and the magnetic field shifts from $h_{0}$ to $h\coloneqq h_{0}-\omega_{0}+\eta N_\text{c}$.
This equivalent Hamiltonian (\ref{CentralSpinContext}) is also known as the generalized Hepp-Coleman model~\cite{Sun06} and we will use it to simplify some calculations.
We emphasize that it makes sense for the simplification that our theoretical starting point will be the effective Hamiltonian~(\ref{H_eff}) without any dissipation in the strongly coupled cavity-QED systems~\cite{PhysRevLett.130.173601} and superconducting circuits~\cite{WOS:000257665300034,PhysRevLett.128.123602}, where the atomic spontaneous emission $\gamma_e$ and the cavity dissipation $\gamma_d$~\cite{WOS:000243867300038,WOS:000322086100035,PhysRevA.71.013817,PhysRevA.67.033806} could be suppressed.

To demonstrate our effective Hamiltonian is a valid approximation, we calculate the expectation $\left\langle M\left(t\right)\right\rangle $ with respect to the final state of the spins.
Here, the observable is $J_{\zeta}=J_{x}\cos\zeta+J_{y}\sin\zeta$, and the initial state $\left|\Psi\right\rangle =\left|\alpha\right\rangle \otimes \left|\mu\right\rangle $ is a product state of the coherent state $\left|\alpha\right\rangle $ and spin-coherent state 
\begin{equation}
\left|\mu\right\rangle  =  \frac{\!\exp\left(\!\mu J_{-}\!\right)\!}{\!\left(\!1\!+\!\left|\mu\right|^{2}\!\right)^{j}}\left|0\right\rangle =\!\frac{1}{\!\left(\!1\!+\!\left|\mu\right|^{2}\!\right)^{j}}\!\sum_{p=0}^{2j}\sqrt{\frac{\left(2j\right)!}{p!\!\left(2j\!-\!p\right)!}}\mu^{p}\!\left|p\right\rangle ,
\end{equation}
where $\left|j,j\right\rangle =\left|0\right\rangle $ is the eigenstate of $J_{z}$ with eigenvalue $j=N_{\text{b}}/2$, and the phase parameter $\mu$ an be parameterized by angles $\left(\theta,\phi\right)$ via the stereographic projection $\mu=e^{i\phi}\tan\frac{\theta}{2}$ and $\phi\in\left[0,2\pi\right)$.

\begin{figure}[t]
	\centering
	\begin{minipage}{1\linewidth}
		\centering
		\begin{overpic}[width=\linewidth]{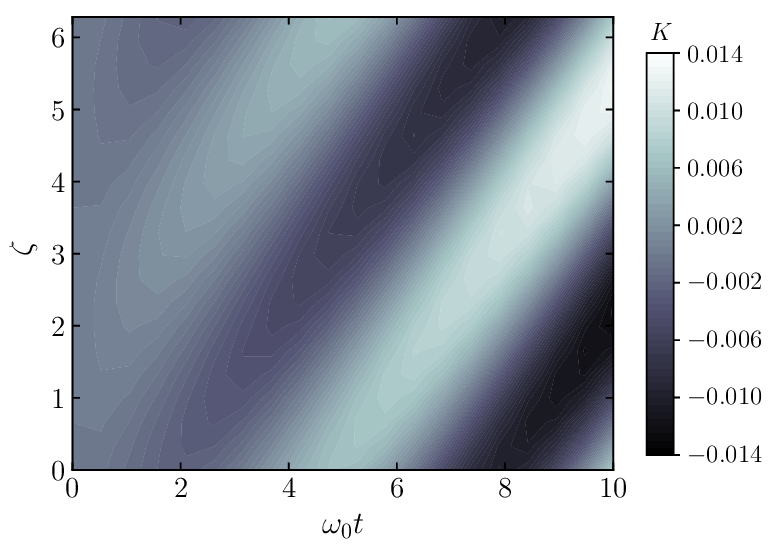}
		\end{overpic}
	\end{minipage}
	\caption{(Color online)  
		Density plot of the difference ratio $K$ as a function of the rescaled time $\omega_0 t$ and arbitrary rotation angle $\zeta$, with the spin number $N_\text{b}=4$, the average photon number $\bar{n}=40$, the original atomic transition frequency $\omega_0/\left(2\pi\right)=6.9$ GHz, the cavity frequency $\omega_{\text{a}}/\left(2\pi\right)=6.89$ GHz, the Rabi frequency $g/\left(2\pi\right)=1.05$ MHz, the nearest neighbor interaction $\lambda=1$, the degree of anisotropy $\gamma=1$, the frequency equivalence of the magnetic field $h_{0}=10^{-5}$ Hz, the parameterized angles $\theta=\pi/2$ and $\phi=0$, the rotation angle $\zeta=\pi/6$.
		More details for experimental parameters are seen in reference~\cite{WOS:000243867300038}, and numerical computations are performed using QuTip~\cite{JOHANSSON20121760}.
	}\label{Fig2}
\end{figure}

Furthermore, we introduce the difference ratio 
\begin{equation}
K=\frac{2\left[\left\langle J_{\zeta}\left(t\right)\right\rangle _{\text{eff}}-\left\langle J_{\zeta}\left(t\right)\right\rangle _{\text{ori}}\right]}{N_\text{b}}
\end{equation}
to quantify the difference of the expectation between the original and effective Hamiltonians for time $t$ and an arbitrary rotation angle $\zeta$.
As seen in \cref{Fig2}, one could find that the difference ratio ($\left|K\right|_{\text{max}}\approx1.28\%$) is acceptably small in our parameter region, which is mostly caused by the validity of $N_\text{b}\ll\bar{n}$, $\Delta\gg g\sqrt{N_{\text{b}}}$ and the finite Hilbert space of the field.
Thus, one could know that the result of the effective Hamiltonian exactly corresponds with the result of the original Hamiltonian, which means the effective Hamiltonian can fully describe the system.


\section{Reduced density matrix of the central spin system}\label{Sec.III}

Here, we derive the evolution of the reduced density matrix for the cavity, with which one could obtain the time evolution of observables and other quantities about the cavity (or, equivalently, central spins).

Our initial density matrix is easy-to-implement:
\begin{equation}
\rho\left(0\right)=\rho_{\text{c}}\left(0\right)\otimes\rho_{\text{b}}\left(0\right),
\end{equation}
where $\rho_{\text{c}}\left(0\right)=\left|\vartheta,\varphi\right\rangle \left\langle \vartheta,\varphi\right|$ and $\rho_{\text{b}}\left(0\right)=e^{-\beta H_0(h)}/Z(\beta,h)=\sum_{i}{p_i\left|\psi_i\right\rangle\left\langle\psi_i\right|}$ are given by the unentangled initial state of the central spins and the thermal state of the bath spins, $Z\left(\beta,h\right)=\text{Tr}\left[\rme^{-\beta H_0\left(h\right)}\right]=\prod_k 2\cosh \left(\beta\Lambda_k/2\right)$ is the partition function, $p_i$ and $\left|\psi_i\right\rangle$ are the $i$-th eigenvalue and eigenstate of $H_0(h)$.
To avoid the ambiguity between the spin-coherent states of the spins in the TC model and the bath spins, we denote the latter as $\left|\vartheta,\varphi\right\rangle$, which can be written in terms of the Dicke states $\left|n\right\rangle $, i.e., $\left|\vartheta,\varphi\right\rangle =\sum_{n=-N_\text{c}/2}^{N_\text{c}/2}{c_{n}\left|n\right\rangle }$ with $S_{z}\left|n\right\rangle =n\left|n\right\rangle $ and $c_{n}=\left|\cos\left(\vartheta/2\right)\right|^{N_\text{c}}\tan^{N_\text{c}/2+n}\left(\vartheta/2\right)$ $\times\rme ^{-\rmi \left(N_\text{c}/2+n\right)\varphi}\sqrt{N_\text{c}!/\left[\left(N_\text{c}/2+n\right)!\left(N_\text{c}/2-n\right)!\right]}$.

Taking advantage of the mapping between the cavity-QED and the central spin system~(\ref{CentralSpinContext}), we  obtain the evolution of the reduced density matrix for the cavity:
\begin{equation}
\rho_{\text{c}}\left(t\right)
=\!\sum_{i,j}{\!c_i c_j^{*}\!\left|i\right\rangle\!\left\langle j\right|\!\mathrm{Tr}_\text{b}\!\left[\rme^{\rmi N_\text{b}tH_j}\rme^{-\rmi N_\text{b}tH_i}\frac{\rme^{-\beta H_0\left(h\right)}}{Z\left(\beta,h\right)}\!\right]},\label{rho_b}
\end{equation}
where $H_{m}=H_0\left(h+2\eta m\right)$ ($m=i,j$) is the effective Hamiltonian.
In fact, the $XY$ Hamiltonian can be diagonalized as $H_0\left(h\right)=\sum_k{\Lambda_k\left(b_k^\dagger b_k-1/2\right)}$ by employing the Jordan-Wigner, Fourier and Bogoliubov transformations~\cite{WOS:000285157400001}, where 
$\Lambda_{k}=\sqrt{\left(h-\lambda \cos k\right)^2+\lambda^2\gamma^2\sin^2k}$ is the excitation energy, $b_k$ and $b_k^\dagger$ are anticommuting fermion operators (see more details in Appendix~B).

We can reconstruct the fermion operators by following the Bogoliubov transformation
\begin{equation}
d_{m,\pm k}=\cos\left(\theta_{m,k}\right)b_{m,\pm k}\mp\rmi \sin\left(\theta_{m,k}\right)\left(b_{m,\mp k}\right)^{\dagger},\label{d}
\end{equation}
where angles $\theta_{m,k}=\left(\mu_{m,k}-\nu_k\right)/2$, $\mu_{m,k}$ and $\nu_k$ are determined by the diagonalization conditions of the effective Hamiltonian $H_{m}$ and the original Hamiltonian $H_0\left(h_{0}\right)$.
From \cref{d}, in the eigenspace $\left\{\prod_{k>0}{\left|\phi_{l}\right\rangle_k};l=0,\pm,2\right\}$ of the original bath spin chain $H_0\left(h_{0}\right)$,
we could gain the reduced density matrix of the central spins (see more details in Appendix~B)
\begin{widetext}
\begin{align}
\rho_{\text{c}}\left(t\right)
&=\sum_{i,j}\bigg\{\frac{c_i c_j^{*}}{Z\left(\beta,h\right)}\left|i\right\rangle\left\langle j\right|\prod_{k>0}\left[2+2\cosh\left(\beta \Lambda_k\right)C\left(\theta_{i,k},E_{i,k}t\right)C\left(\theta_{j,k},E_{j,k}t\right)\right.\nonumber\\
&\quad \left.+\rme^{\beta \Lambda_k}A\left(\theta_{j,k},E_{j,k}t\right)A^{*}\left(\theta_{i,k},E_{i,k}t\right)+\rme^{-\beta \Lambda_k}B\left(\theta_{j,k},E_{j,k}t\right)B^{*}\left(\theta_{i,k},E_{i,k}t\right)
\right]\bigg\},\label{Final StateContext}
\end{align}
\end{widetext}
where the coefficients are defined as $A\left(X,Y\right)=\rme^{-\rmi Y}+2\rmi \sin^2X\sin Y$, $B\left(X,Y\right)=\rme^{-\rmi Y}+2\rmi \cos^2X\sin Y$, $C\left(X,Y\right)=\sin2X\sin Y$ and $X$,$Y$ are real variables, and $E_{m,k}=E_{m,-k}=\sqrt{\left(h+2\eta m-\lambda\cos k\right)^2+\lambda^2\gamma^2\sin^2 k}$ are  the excitation energy.
%


%

\begin{figure}[htpb]
	\centering
	\begin{minipage}{0.49\linewidth}
		\centering
		\begin{overpic}[width=\linewidth]{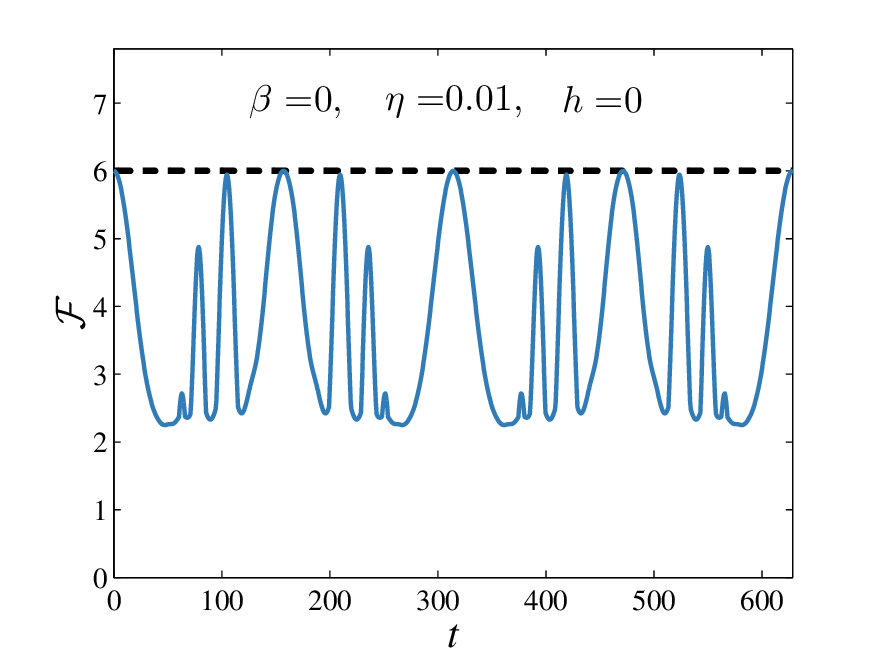}
			\put(0,65){(a)}
		\end{overpic}
	\end{minipage}
	\begin{minipage}{0.49\linewidth}
		\centering
		\begin{overpic}[width=\linewidth]{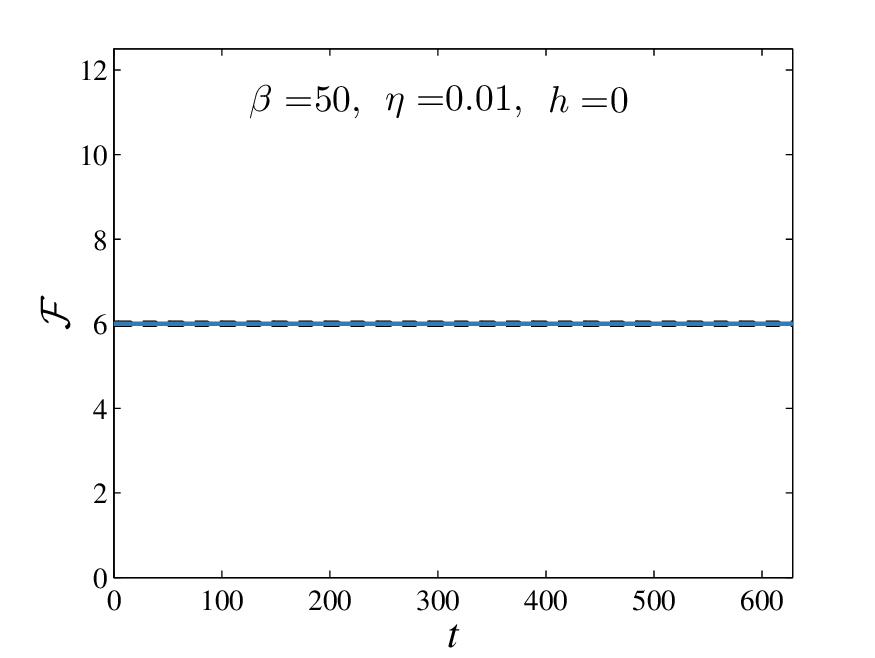}
			\put(0,65){(b)}
		\end{overpic}
	\end{minipage}\\
	\begin{minipage}{0.49\linewidth}
		\centering
		\begin{overpic}[width=\linewidth]{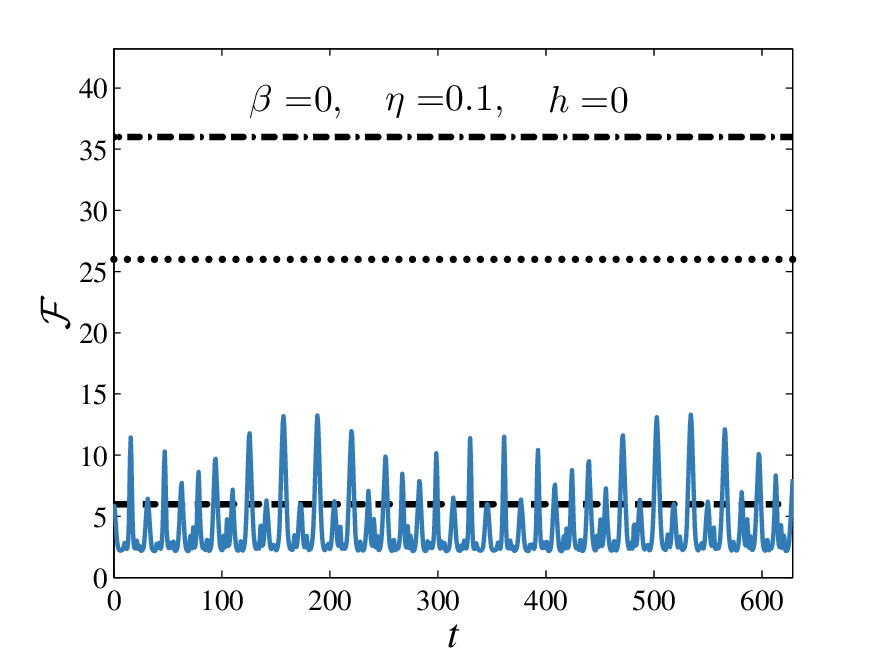}
			\put(0,65){(c)}
		\end{overpic}
	\end{minipage}
	\begin{minipage}{0.49\linewidth}
		\centering
		\begin{overpic}[width=\linewidth]{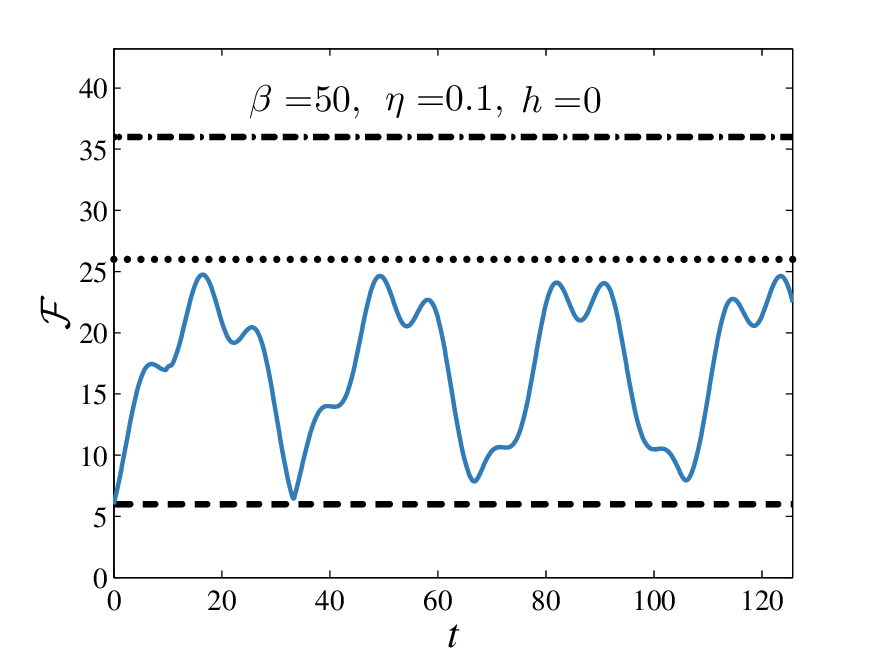}
			\put(0,65){(d)}
		\end{overpic}
	\end{minipage}\\
	\begin{minipage}{0.49\linewidth}
		\centering
		\begin{overpic}[width=\linewidth]{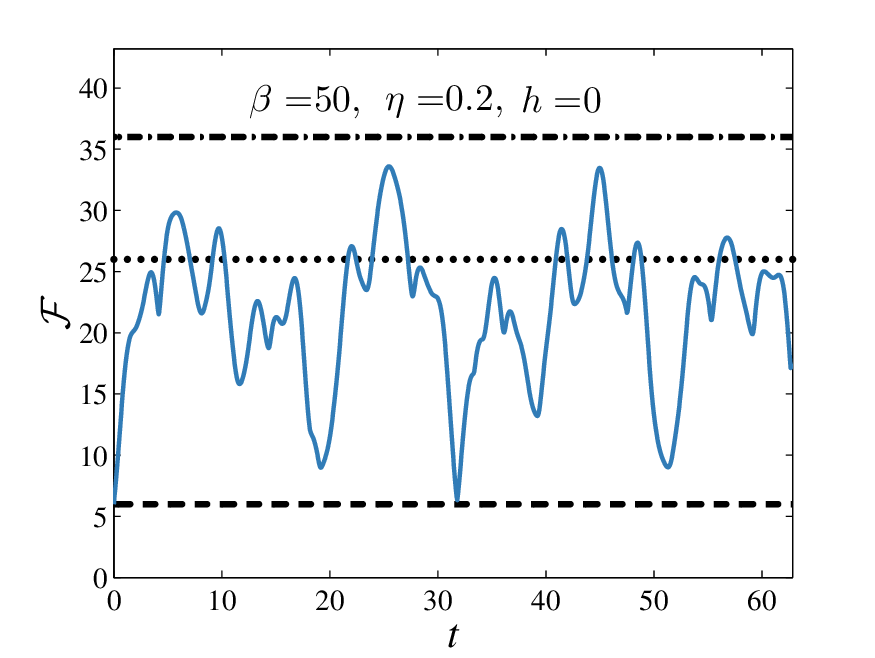}
			\put(0,65){(e)}
		\end{overpic}
	\end{minipage}
	\begin{minipage}{0.49\linewidth}
		\centering
		\begin{overpic}[width=\linewidth]{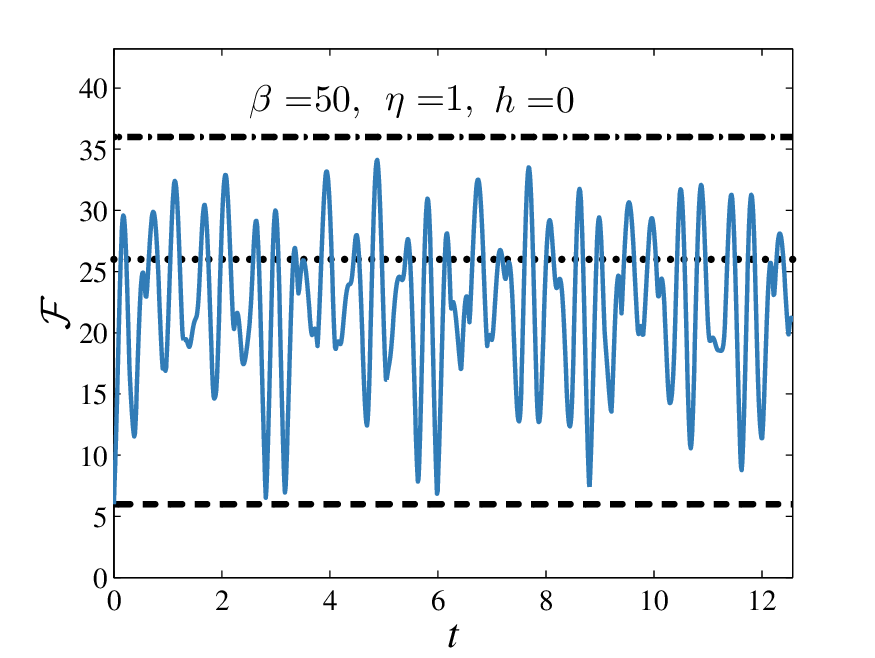}
			\put(0,65){(f)}
		\end{overpic}
	\end{minipage}
	\caption{(Color online) 
		Dynamical evolutions of multipartite entanglement of central spins for the $XX$ chain with the different inverse temperatures $\beta$ and coupling constants $\eta$.
		We set $\gamma=0$, $N_{\text{c}}=6$, $N_{\text{b}}=10$, $\vartheta=\pi/2$, and $\varphi=0$.
		We appoint that dashed line, dots, and dot-dash line indicate the multipartite entanglement criterion for two-partite ($N_{\text{c}}$), $N_{\text{c}}$-partite ($\left(N_\text{c}-1\right)^2+1$) and max multipartite ($N_{\text{c}}^2$) entanglements.
	}\label{Fig3}
\end{figure}

\section{Multiparticle entanglement of the central spin system}\label{Sec.IV}

Here, we introduce the multipartite entanglement based on the linear operator metrological detection to quantify the induced dynamical entanglement among the central spins.

For a mixed state $\rho=\sum_{m}{p_{m}\left|m\right\rangle\left\langle m\right|}$ and a collective operator $S$, the QFI, defined as~\cite{PhysRevLett.72.3439}
\begin{align}
\mathcal{F}\left[\rho,H\right]=2\sum_{m,n=1}^{2^{N_\text{c}}}{\frac{\left(p_m-p_n\right)^2}{p_m+p_n}\left\langle m\right|S^{\alpha}\left|n\right\rangle\left\langle n\right|S^{\alpha^{\prime}}\left|m\right\rangle},
\end{align}
gives the maximum precision for the estimated parameter $\xi$ by optimizing over the positive-operator valued measures, where evolution state is $\rho\left(t\right)=\rme^{-\rmi \xi H}\rho\rme^{\rmi \xi H}$.

\begin{figure}[t]
	\centering
	\begin{minipage}{0.49\linewidth}
		\centering
		\begin{overpic}[width=\linewidth]{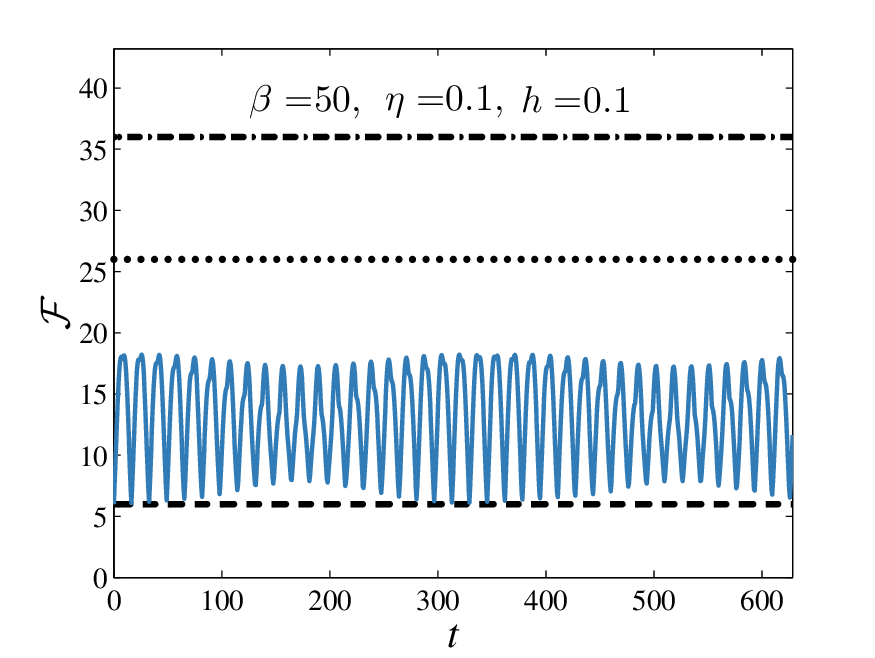}
			\put(0,65){(a)}
		\end{overpic}
	\end{minipage}
	\begin{minipage}{0.49\linewidth}
		\centering
		\begin{overpic}[width=\linewidth]{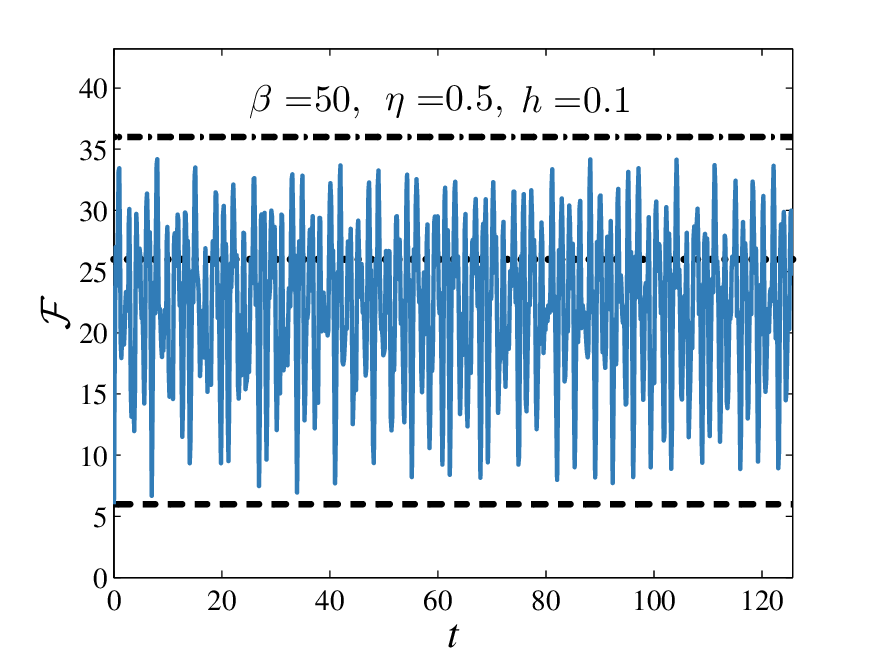}
			\put(0,65){(b)}
		\end{overpic}
	\end{minipage}\\
	\begin{minipage}{0.49\linewidth}
		\centering
		\begin{overpic}[width=\linewidth]{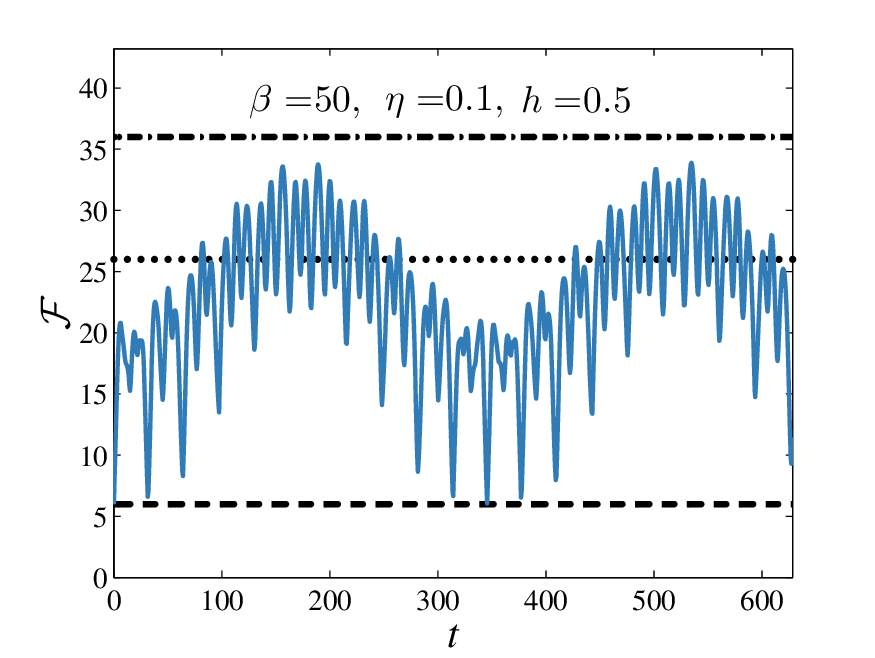}
			\put(0,65){(c)}
		\end{overpic}
	\end{minipage}
	\begin{minipage}{0.49\linewidth}
		\centering
		\begin{overpic}[width=\linewidth]{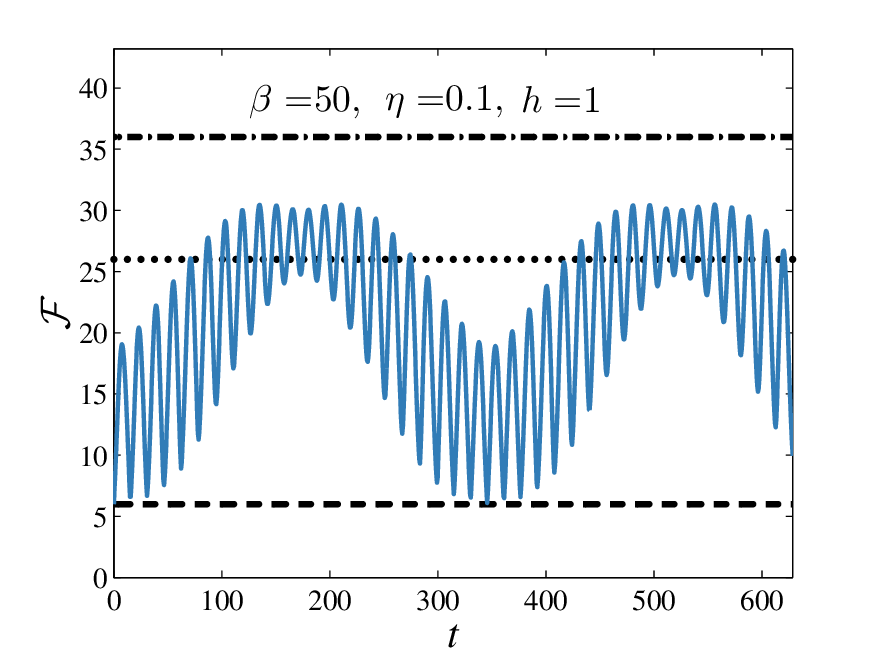}
			\put(0,65){(d)}
		\end{overpic}
	\end{minipage}
	\caption{(Color online) 
		Dynamical evolutions of multipartite entanglement of central spins for the $XX$ chain with the different coupling constants $\eta$ and magnetic fields $h$.
		We set $\gamma=0$, $\beta=50$, $N_{\text{c}}=6$, $N_{\text{b}}=10$, $\vartheta=\pi/2$, and $\varphi=0$.
	}\label{Fig4}
\end{figure}

For a mixed state $\rho_{\text{c}}\left(t\right)=\sum_{j=1}^{2^{N_\text{c}}}{p_j\left|j\right\rangle\left\langle j\right|}$ of the central spins, the multipartite entanglement is given by~\cite{PhysRevA.85.022321,PhysRevA.85.022322}
\begin{align}
\mathcal{F}\!\left[\rho_{\text{c}}\!\left(\!t\!\right)\!,\!\vec{n}\!\cdot\!\vec{S}\right]\!\equiv\!\max_{\left|\vec{n}\right|=1}{\!\mathcal{F}\!\left[\rho_{\text{c}}\!\left(\!t\!\right)\!,\!\vec{n}\!\cdot\!\vec{S}\right]}\!=\!\max_{\left|\vec{n}\right|=1}\vec{n}^{T}\Gamma\vec{n}=p_{\max}\left(\Gamma\right),
\end{align}
where $\vec{n}$ is a normalized vector on the Bloch sphere, the matrix 
\begin{align}
\Gamma_{\alpha \alpha^{\prime}}=2\sum_{m,n=1}^{2^{N_\text{c}}}{\frac{\left(p_m-p_n\right)^2}{p_m+p_n}\left\langle m\right|S^{\alpha}\left|n\right\rangle\left\langle n\right|S^{\alpha^{\prime}}\left|m\right\rangle}
\end{align}
carries all the information needed to maximize $\mathcal{F}\left[\rho_{\text{c}}\left(t\right),\vec{n}\cdot\vec{S}\right]$ for an arbitrary direction $\vec{n}$, $p_{\max}\left(\Gamma\right)$ is the maximal eigenvalue of the matrix $\Gamma$ and $\alpha,\alpha^{\prime}=x,y,z$.
The central spins is at least $\left(l+1\right)$-particle entangled if 
\begin{align}
\mathcal{F}\left[\rho_{\text{c}}\left(t\right),\vec{n}\cdot\vec{S}\right]>sl^2+r^2,\label{F}
\end{align} 
where $s=\lfloor N_\text{c}/l\rfloor$ is the largest integer smaller than or equal to $N_\text{c}/l$ and $r=N_\text{c}-sl$.
The genuine multipartite entanglement, implying the $N_\text{c}$-particle entanglement $\mathcal{F}\left[\rho_{\text{c}}\left(t\right),\vec{n}\cdot\vec{S}\right]>\left(N_\text{c}-1\right)^2+1$, is indispensable to the maximal phase sensitivity in quantum interferometry.

It is worth noting that, in principle, we could set $N_{\text{c}}\gg \langle a^\dag a\rangle\gg N_{\text{b}}$ to meet the requirements of mapping the TC model to the central spin model.
Still, the numerical feasibility of the demonstration is almost impossible.
Thus, for convenience, we illustrate the dynamical multipartite entanglement of the central spin model with $N_{\text{c}}=6$ and $N_{\text{b}}=10$.

\begin{figure}[t]
	\centering
	\begin{minipage}{1\linewidth}
		\centering
		\begin{overpic}[width=1.2\linewidth]{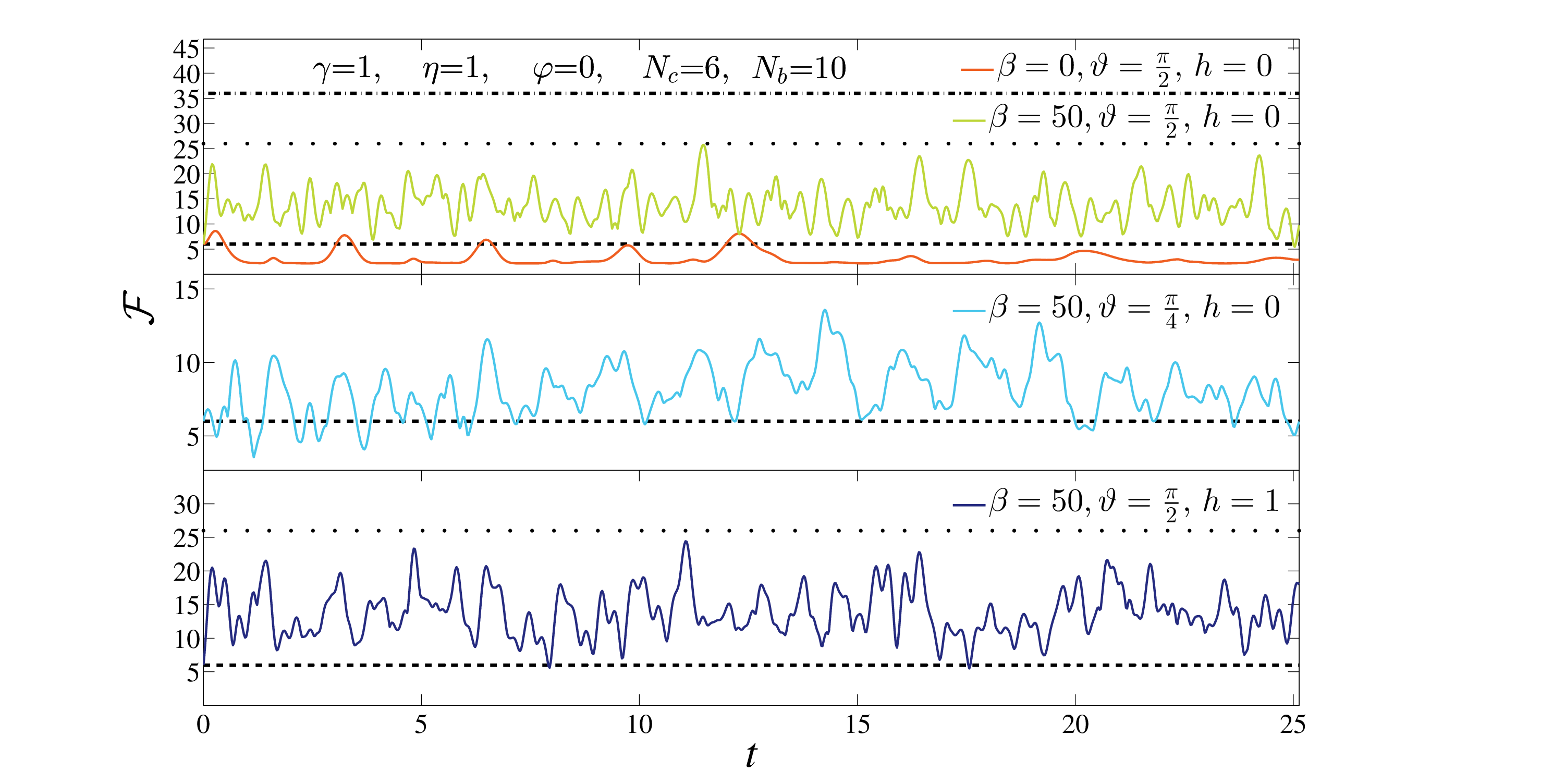}
			\put(15,45){(a)}
			\put(15,30){(b)}
			\put(15,17.5){(c)}
		\end{overpic}
	\end{minipage}
	\caption{(Color online) 
		Dynamical evolutions of multipartite entanglement of central spins for the Ising chain with the different inverse temperatures $\beta$, angles $\vartheta$ of a Dicke state, and magnetic fields $h$.
		We set $\gamma=1$, $\eta=1$, $\varphi=0$, $N_{\text{c}}=6$, and $N_{\text{b}}=10$.
	}\label{Fig5}
\end{figure}

%


\begin{figure}[htpb]
	\centering
	\begin{minipage}{0.49\linewidth}
		\centering
		\begin{overpic}[width=\linewidth]{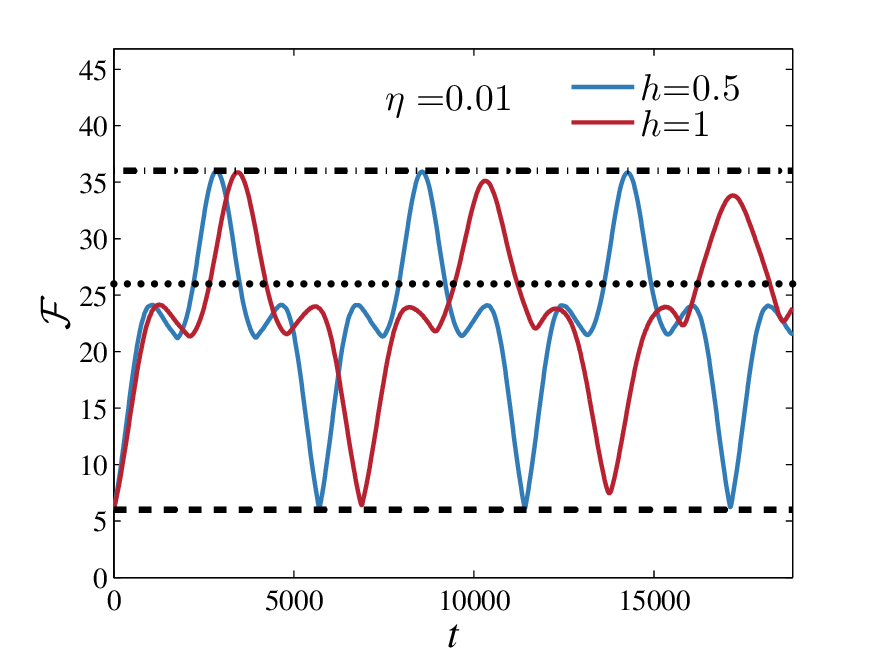}
			\put(0,65){(a)}
		\end{overpic}
	\end{minipage}
	\begin{minipage}{0.49\linewidth}
		\centering
		\begin{overpic}[width=\linewidth]{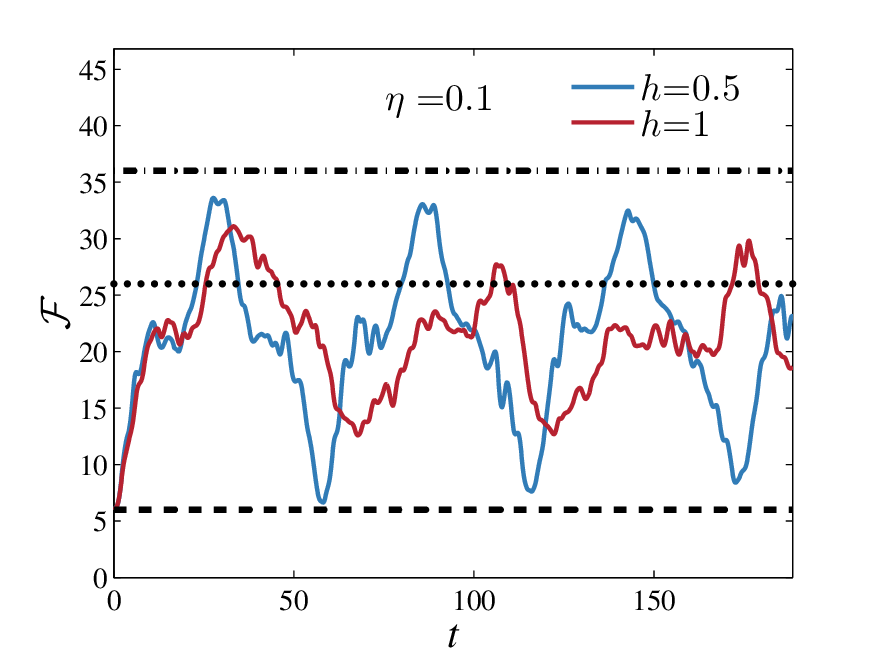}
			\put(0,65){(b)}
		\end{overpic}
	\end{minipage}\\
	\begin{minipage}{0.49\linewidth}
		\centering
		\begin{overpic}[width=\linewidth]{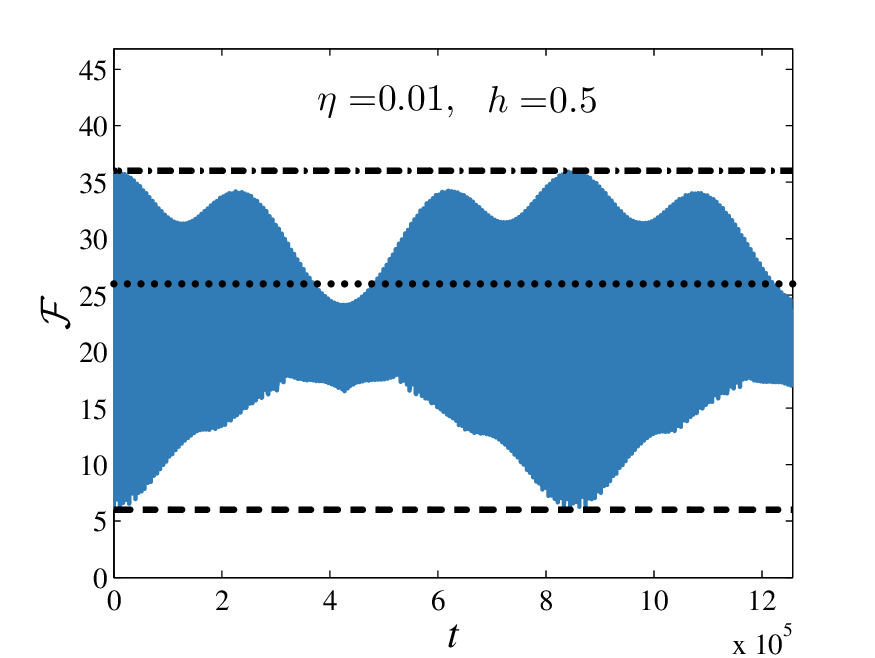}
			\put(0,65){(c)}
		\end{overpic}
	\end{minipage}
	\begin{minipage}{0.49\linewidth}
		\centering
		\begin{overpic}[width=\linewidth]{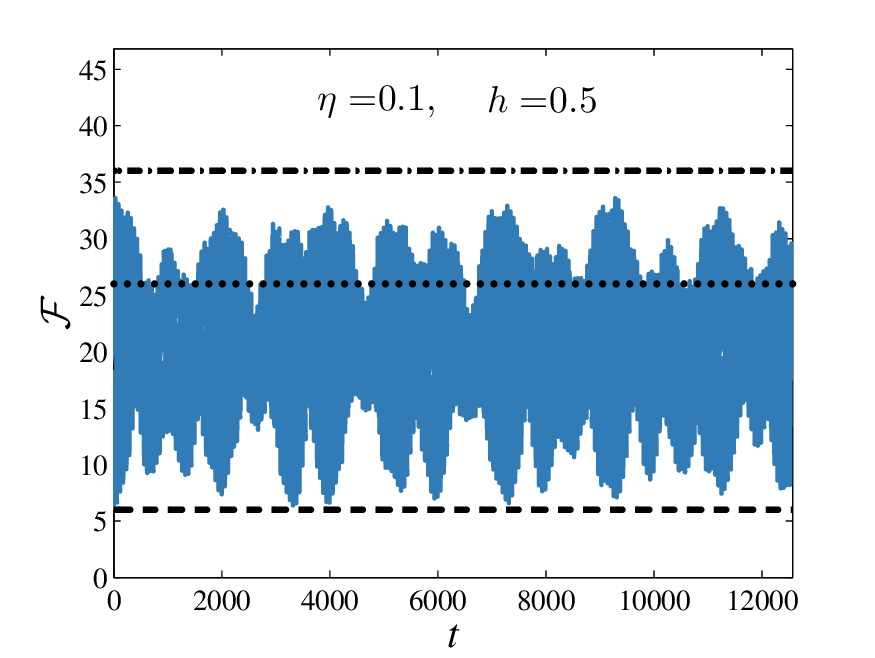}
			\put(0,65){(d)}
		\end{overpic}
	\end{minipage}\\
	\begin{minipage}{0.49\linewidth}
		\centering
		\begin{overpic}[width=\linewidth]{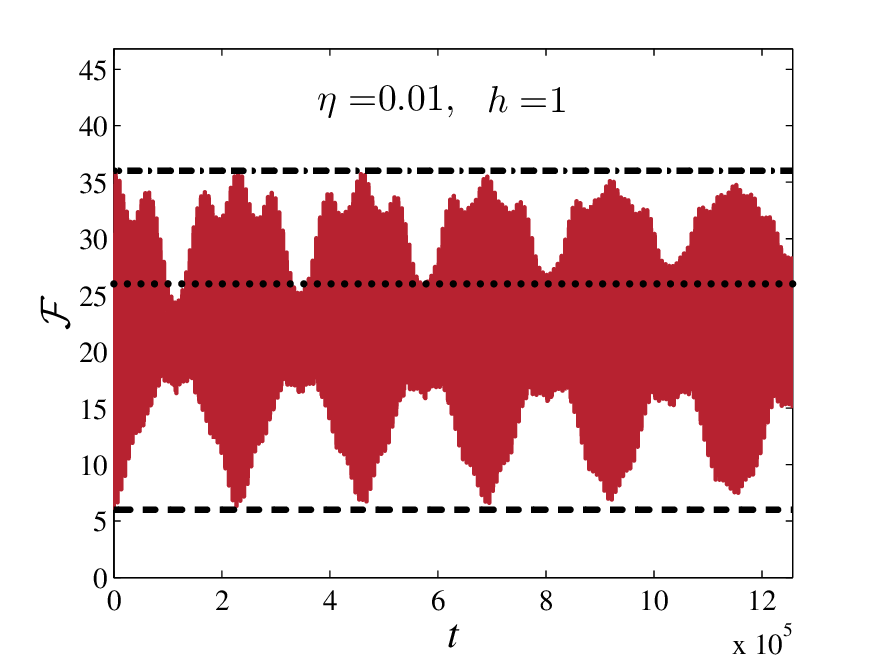}
			\put(0,65){(e)}
		\end{overpic}
	\end{minipage}
	\begin{minipage}{0.49\linewidth}
		\centering
		\begin{overpic}[width=\linewidth]{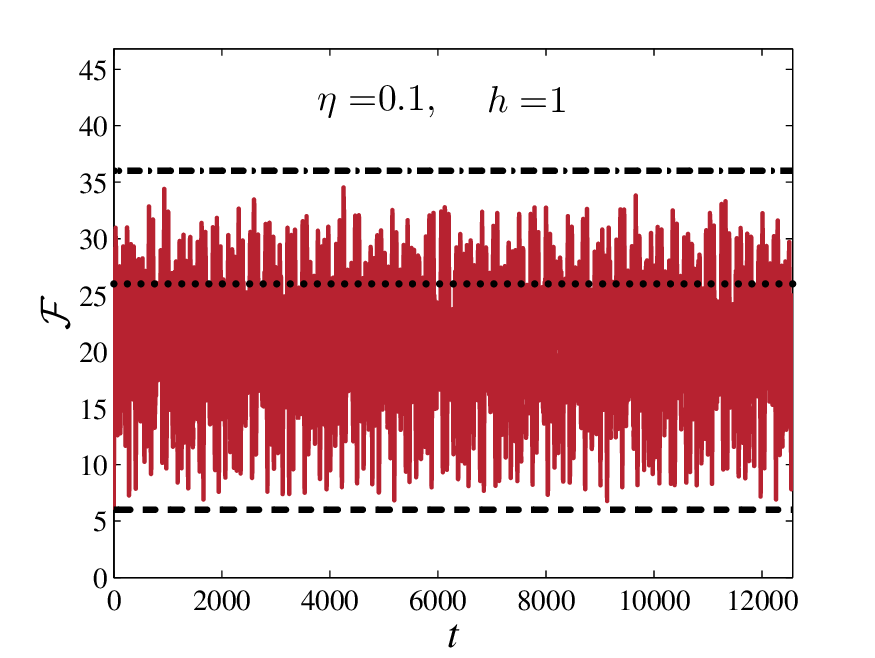}
			\put(0,65){(f)}
		\end{overpic}
	\end{minipage}
	\caption{(Color online) 
		Short and long-dynamical evolutions of multipartite entanglement of central spins for the Ising chain with the different coupling constants $\eta$ and magnetic fields $h$:
		(a) $\eta=0.01$ and $h=0.5,1$; (b) $\eta=0.1$ and $h=0.5,1$.
		(c) , (e) and (d), (f) are the long-dynamical evolutions of multipartite entanglement corresponding to (a) and (b).
		We set $\gamma=1$, $\beta=50$, $\vartheta=\pi/2$, $\varphi=0$, $N_{\text{c}}=6$, and $N_{\text{b}}=10$.
	}\label{Fig6}
\end{figure}

We first consider the case of the $XX$ model ($\gamma=0$).
Figure~\ref{Fig3} shows multipartite entanglement of central spins with the $XX$ chain with the inverse temperature $\beta$, and the coupling constant $\eta$.
As seen in Figs.~\ref{Fig3} (a) and \ref{Fig3}(b), the low temperature (large $\beta$) could not directly induce the multipartite entanglement between central spins, however, it might promote the value of the entanglement.
Multipartite entanglement is witnessed by the criterion~(\ref{F}), i.e., $\mathcal{F}>N_\text{c}=6$, with a stronger coupling constant even when the temperature is high [Fig.~\ref{Fig3} (c)] and presents a considerable value when the temperature is low [Fig.~\ref{Fig3} (d)].
As shown in Figs.~\ref{Fig3}(c)-(f), one could find that the coupling constant between central spins and bath spins plays a vital role in the emergence of multipartite entanglement.
Concretely, from Figs.~\ref{Fig3}(e) and \ref{Fig3}(f),  a genuine multipartite entanglement, i.e., $\mathcal{F}>\left(N_\text{c}-1\right)^2+1=26$, emerges with a strong coupling constant.

When it comes to the role of the magnetic field in the emergence of multipartite entanglement, we could observe an envelope of dynamical multipartite entanglement.
Concretely, as seen in Fig.~\ref{Fig4}, both the large magnetic field and coupling constant modulate the period and amplitude of multipartite entanglement for the inverse temperature $\beta=50$ and the $XX$ chain.

For the Ising model ($\gamma=1$), the behavior of multipartite entanglement is analogous to the corresponding one of the $XX$ model with the strong coupling constant, as shown in Fig.~\ref{Fig5}.

Figure~\ref{Fig6} illustrates the short and long-dynamical evolutions of multipartite entanglement for the Ising model.
When the coupling constant is weak, one could find that different magnetic fields preserve the analogous waveform for the short-dynamics, while, only the period and amplitude of multipartite entanglement are modulated [Fig.~\ref{Fig6}(a)].
As seen in Figs.~\ref{Fig6}(c) and \ref{Fig6}(e), the long-dynamical evolutions of multipartite entanglement present stable envelopes for arbitrary magnetic fields.
However, a stronger coupling constant ruins the stable behavior of multipartite entanglement [Figs.~\ref{Fig6}(b), \ref{Fig6}(d) and \ref{Fig6}(f)].

\begin{figure}[t]
	\centering
	\begin{minipage}{\linewidth}
		\centering
		\begin{overpic}[width=\linewidth]{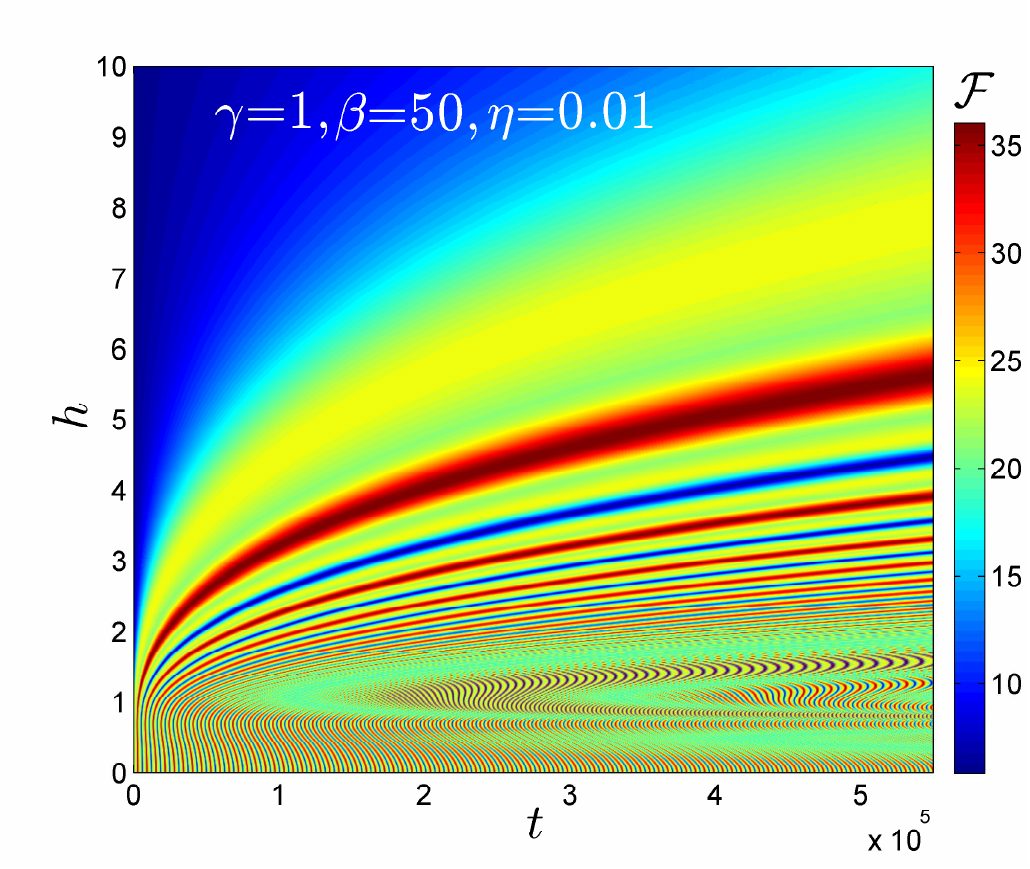}
						\put(0,80){(a)}
		\end{overpic}
	\end{minipage}
\\
	\begin{minipage}{\linewidth}
		\begin{overpic}[width=\linewidth]{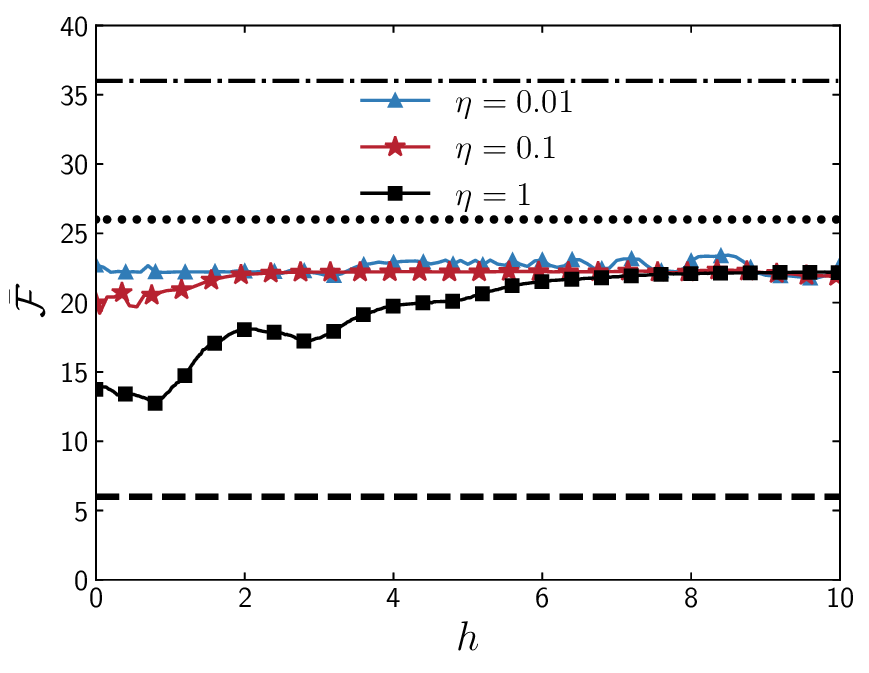}
						\put(0,77.5){(b)}
		\end{overpic}
	\end{minipage}
	\caption{(Color online) 
		Effect of the magnetic field on multipartite entanglement of central spins for the Ising chain.
		(a) Density plot of multipartite entanglement as a function of time $t$ and the magnetic field $h$.
		(b) Time-averaged dynamical entanglement (averaging time up to $10^{4}\times2\pi/\eta$) versus the magnetic field.
		The blue triangles, red stars, and black squares denote the different coupling constant $\eta=0.01,$ 0.1, and 1, respectively.
		We set $\gamma=1$, $\beta=50$, $N_{\text{c}}=6$, $N_{\text{b}}=10$, $\vartheta=\pi/2$, and $\phi=0$.
	}\label{Fig7}
\end{figure}

To elucidate the effect of the magnetic field, we show the density plot of multipartite entanglement.
In Fig.~\ref{Fig7}(a), the genuine multipartite entanglements correspond to a series of magnetic fields, e.g., the red regions.
Moreover, we illustrate the effect of the magnetic field on the time-averaged dynamical entanglement (averaging time up to $10^{4}\times2\pi/\eta$) for the Ising chain.
As shown in Fig.~\ref{Fig7}(b), multipartite entanglement could be witnessed in the central spin system with the Ising model, since the significant time-averaged QFI $\bar{\mathcal{F}}$ is observed.

\section{Conclusion}\label{Sec.V}

In summary, we have presented the dynamical multipartite entanglement, which is in terms of the QFI, of a generalized TC model introducing the $XY$ spin interaction.
By theoretically deriving an effective description of our model, we circumvented the challenge that our model cannot be solved exactly, and numerically verified its validity. 
The generalized TC model can be mapped onto the central spin model by the HP transformation.
Besides, the reduced density matrix of the central spins is also provided, which is the prerequisite for quantifying multipartite entanglement.
We also investigated the impacts of the temperature, the coupling constant, and the magnetic field on the dynamical multipartite entanglement in the central spin model, where the central spin is initially unentangled.
Concretely, we found that the strong coupling and low temperature are necessary conditions for achieving a genuine multipartite entanglement.
The former governs the maximum entanglement, while the latter plays a crucial role in stabilizing the entanglement.
Our results shed light on the deep link between the TC model and the central spin model, and provide deeper insights into dynamical multipartite entanglement in them.


\section*{Acknowledgments}
We thank Hai-Long Shi for discussions.
This work is supported by the National Natural Science Foundation of China Grant No.~12247158, the ``Wuhan Talent'' (Outstanding Young Talents), and the Postdoctoral Innovative Research Post in Hubei Province.
\begin{appendix}
\setcounter{section}{1}
\setcounter{equation}{0}

\section*{Appendix A.~Effective Hamiltonian of the cavity-QED system}\label{Appendix_A}
Here, we derive the effective Hamiltonian of the cavity-QED system.
Utilizing the unitary transformation $U=\exp\left\{-\rmi \left[\omega_0 J_z+\omega_{\text{a}} a^\dagger a+H_0\left(h_{0}\right)\right]t\right\}$,
the total Hamiltonian in the interaction picture could be written as
\begin{equation}
H=g\left(\rme ^{\rmi \omega_{a}t}a^{\dagger}\rme ^{\rmi \left(\omega_{0}J_{z}+H_{0}\left(h_{0}\right)\right)t}J_{-}\rme ^{-\rmi \left(\omega_{0}J_{z}+H_{0}\left(h_{0}\right)\right)t}+\text{H.c.}\right).
\end{equation}
Under the high atomic frequency condition ($\left|h_{0}-\omega_{0}\right|\gg\lambda/2$), the Hamiltonian can be written as $H=g\left[J_{-}a^{\dagger}\rme ^{-\mathrm{i}\left(\Delta+\delta\right)t}+\text{H.c.}\right]$ with a large effective detuning $\Delta=\omega_{0}-h_{0}-\omega_{a}$ and a small field-irrelative residue $\delta$ ($\Delta\gg\delta\sim\lambda$\ or\ $\lambda\gamma$)
. 
Employing the time-averaged method of the reference~\cite{James2007}, the interaction Hamiltonian $H=\sum_{i=1}^{N_\text{b}}{\left(f_{i}\rme ^{-\mathrm{i}\Delta_{i}t}+\text{H.c.}\right)}$ could be rewritten as the following compact form: 
\begin{align}
H&\approx\sum_{i,j=1}^{N_\text{b}}{\frac{1}{\bar{\Delta}_{ij}}\left[f_{i}^{\dagger},f_{j}\right]\rme ^{\mathrm{i}\left(\Delta_{i}-\Delta_{j}\right)t}}\nonumber\\
&=-\frac{g^{2}}{\left(\Delta+\delta\right)}\left[J_{-}a^{\dagger},J_{+}a\right]\nonumber\\
&\approx\frac{g^{2}}{\Delta}\left(J_{+}J_{-}+2J_{z}a^{\dagger}a\right),\label{H_TAM}
\end{align}
where the operator $f_{i}=g\sigma_{i}^{+}a/2$ and the frequency $\Delta_{i}=\bar{\Delta}_{ij}\equiv\left|\Delta+\delta\right|\approx\left|\Delta\right|\gg1$ are corresponding to our system. 
The fact that the high-frequency contributions disappear from the average and the effective detuning is large, guarantees the validities of the first and second approximations in Eq.~(\ref{H_TAM}).
In the Schr\"{o}dinger picture, we gain the effective Hamiltonian (presented as Eq.~(\ref{H2}) in the main text) 
\begin{equation}
H_{\text{eff}}=H_{0}\left(h_{0}-\omega_{0}\right)+\frac{2g^{2}}{\Delta}J_{z}a^{\dagger}a+\frac{g^{2}}{\Delta}J_{+}J_{-}
\end{equation}
by considering the initial light state $\left|\alpha\right\rangle $ and the photon energy conservation $\left[\omega_{a}a^{\dagger}a,H_{\text{eff}}\right]=0$.
Alternatively, the Fr{\"o}lich-Nakajima transform is performed to obtain the effective Hamiltonian~\cite{MA201189}, and the validity of the approximation in Eq.~\eqref{H2} is guaranteed by the large-detuning condition, i.e., $\Delta\gg g\sqrt{N_{\text{b}}}$.


\setcounter{section}{2}
\setcounter{equation}{0}
\section*{Appendix B.~Reduced density matrix of the central spin system}\label{Appendix_B}

The Hamiltonian of the general anisotropic $XY$ model in a transverse field is given by 
\begin{equation}
H_{0}\left(\!h_{0}\!\right)=\!-\frac{1}{2}\!\sum_{i=1}^{N_\text{b}}\!{\!\left\{\!\frac{\lambda}{2}\!\left[\left(\!1\!+\!\gamma\right)\sigma^x_i\sigma^x_{i+1}\!+\!\left(\!1\!-\!\gamma\right)\sigma^y_i\sigma^y_{i+1}\!\right]\!+\!h_{0}\sigma^z_i\!\right\}}.
\end{equation}
Here $\lambda$ is the nearest neighbor interaction, $\sigma_{i}^{\alpha}$ the Pauli matrix ($\alpha=x,y,z$) on site $i$, $N_\text{b}$ the number of sites, $\gamma$ the degree of anisotropy, and $h$ a transverse field.

By applying the Jordan-Wigner, Fourier and Bogoliubov transformations~\cite{WOS:000285157400001}, the Hamiltonian can be rewriten as 
\begin{equation}
H_{0}\left(h_{0}\right)=\sum_{k}{\Lambda_{k}\left(b_{k}^{\dagger}b_{k}-\frac{1}{2}\right)}.\label{DiaH}
\end{equation}
The anticommuting fermion operators $b_{k}$ and $b_{k}^{\dagger}$ satisfy the relations $\left\{ b_{i}^{\dagger},b_{j}\right\} =\delta_{i,j}$ and $\left\{ b_{i},b_{j}\right\} =\left\{ b_{i}^{\dagger},b_{j}^{\dagger}\right\}=\left[b_{i}^{\dagger}b_{i},b_{j}^{\dagger}b_{j}\right]=0$
, the excitation energy is 
\begin{equation}
\Lambda_{k}=\sqrt{\left(h_{0}-\lambda\cos k\right)^{2}+\lambda^{2}\gamma^{2}\sin^{2}k},
\end{equation}
and the ground state is 
\begin{equation}
\left|\phi\right\rangle =\prod_{k}\left|\phi_{0}\right\rangle _{k}=\prod_{k>0}{\cos\frac{\nu_{k}}{2}\left|0\right\rangle _{k}\left|0\right\rangle _{-k}+\mathrm{i}\sin\frac{\nu_{k}}{2}\left|1\right\rangle _{k}\left|1\right\rangle _{-k}},
\end{equation}
where $k=2\pi m/N_\text{b}$ ($m=-\frac{N_\text{b}}{2}+1,\dots,\frac{N_\text{b}}{2}$), $\left|0\right\rangle _{k}$ ($\left|1\right\rangle _{k}$) is the state for $k$th particle, and the coefficients are determined by $\sin\nu_{k}=\lambda\gamma\sin k/\Lambda_{k},\cos\nu_{k}=\left(\lambda\cos k-h_{0}\right)/\Lambda_{k}.$
The partition function is 
\begin{equation}
Z\!\left(\!\beta,h_{0}\!\right)\!=\!\text{Tr}\!\left(\!\rme ^{\!-\!\beta H_{0}\left(\!h_{0}\!\right)\!}\right)\!=\!\prod_{k}{\!\sum_{n_{k}}{\!\rme ^{\!-\!\beta\Lambda_{k}\left(\!n_{k}\!-\!\frac{1}{2}\!\right)}}}\!=\!\prod_{k}\!2\!\cosh\!\left(\beta\Lambda_{k}\!/2\right)\!,
\end{equation}
$\beta$ is the inverse temperature and the QPT points are $h_{\text{c}}=1$,
$\gamma_{\text{c}}=0$ when $h_{0}\in\left[-1,1\right]$. 

The total Hamiltonian of the system (shown as Eq.~(\ref{CentralSpinContext}) in main text when the magnetic field shifts from $h_{0}$ to $h\coloneqq h_{0}-\omega_{0}+\eta N_\text{c}$) is 
\begin{equation}
H=H_{0}\left(h\right)\otimes I_{2N_\text{c}\times2N_\text{c}}+2\eta H_{1}\otimes S_{z},\label{Orig.Centr.Spin}
\end{equation}
where $H_{0}\left(h\right)$ is the Hamiltonian of the central, $\eta H_{1}\otimes S_{z}$ is the Hamiltonian of the spin-bath interaction, $\eta$ is a coupling constant, $H_{1}=-\sum_{j=1}^{N_\text{b}}{\sigma_{j}^{z}}/2$ acts as the random field for the central spin, $S_{z}\!=\!\sum_{j=1}^{N_\text{c}}{\!\sigma_{j}^{z}/2}\!=\!\sum_{j=1}^{N_\text{c}}{\!\left(\!\left|e\right\rangle _{jj}\!\left\langle e\right|\!-\!\left|g\right\rangle _{jj}\!\left\langle g\right|\!\right)\!/2}$ is the total central spin operator, and $I_{2N_\text{c}\times2N_\text{c}}$ is the identity operator. The central spin lying in the bath is equally coupled to all the $N_\text{c}$ central spins.
Employing the identical equations $\left|e\right\rangle _{jj}\!\left\langle e\right|=\left(1+\sigma_{j}^{z}\right)/2$ and $\left|g\right\rangle _{jj}\!\left\langle g\right|=\left(1-\sigma_{j}^{z}\right)/2$, the total Hamiltonian of the system can be rewritten as 
\begin{align}
H & =  \left(\frac{H_{0}\left(h\right)}{N_\text{c}}+\eta H_{1}\right)\otimes\sum_{j}^{N_\text{c}}{\left|e\right\rangle _{jj}\!\left\langle e\right|}\nonumber\\
&\quad+\left(\frac{H_{0}\left(h\right)}{N_\text{c}}-\eta H_{1}\right)\otimes\sum_{j}^{N_\text{c}}{\left|g\right\rangle _{jj}\!\left\langle g\right|}\nonumber \\
& =  H_{+}\otimes\left(\frac{N_\text{c}}{2}+S_{z}\right)+H_{-}\otimes\left(\frac{N_\text{c}}{2}-S_{z}\right),\label{CentralSpin}
\end{align}
where $H_{\pm}=H_{0}\left(h\right)/N_\text{c}\pm\eta H_{1}=H_{0}\left(h\pm\eta N_\text{c}\right)/N_\text{c}$ denote the effective Hamiltonians of two slightly different evolution branches. 

We consider the initial states of the central spins and bath spins to be a spin-coherent state $\left|\vartheta,\varphi\right\rangle $ (unentangled) and a thermal state $\left|\psi\right\rangle _{\text{thermal}}$, respectively.
Utilizing the unitary transformation, we can gain the time evolution density matrix of the system 
\begin{equation}
\rho\left(t\right)=U\left(\left|\vartheta,\varphi\right\rangle \left\langle \vartheta,\varphi\right|\otimes\rho_{\text{b}}\left(0\right)\right)U^{\dagger},\label{rho_t}
\end{equation}
where the initial state is $\rho\left(0\right)=\left|\vartheta,\varphi\right\rangle \left\langle \vartheta,\varphi\right|\otimes\rho_{\text{b}}\left(0\right)$, the initial state $\rho_{\text{b}}\left(0\right)=e^{-\beta H_0(h)}/Z(\beta,h)=\sum_{i}{p_i\left|\psi_i\right\rangle\left\langle\psi_i\right|}$ of the bath spin, and the unitary matrix is $U\equiv\exp{\left(-\rmi Ht\right)}$.
Here $p_i$ and $\left|\psi_i\right\rangle$ are the  $i$-th eigenvalue and eigenstate of $H_0(h)$.
By representing the spin-coherent state in terms of the Dicke states, i.e., $\left|\vartheta,\varphi\right\rangle =\sum_{n=-N_\text{c}/2}^{N_\text{c}/2}{c_{n}\left|n\right\rangle }$ with $S_{z}\left|n\right\rangle =n\left|n\right\rangle $ and $c_{n}\!=\!\left|\cos\left(\!\vartheta/2\!\right)\right|^{N_\text{c}}\!\tan^{N_\text{c}/2+n}\left(\!\vartheta/2\!\right)$ $\times\rme ^{-\rmi \left(N_\text{c}/2+n\right)\varphi}\sqrt{N_\text{c}!/\left[\left(N_\text{c}/2+n\right)!\left(N_\text{c}/2-n\right)!\right]}$, the reduced density matrix of the central spins (presented as Eq.~(\ref{rho_b}) in the main text) is 
\begin{widetext}
\begin{align}
\rho_{\text{c}}\left(t\right) & \!= \! \mathrm{Tr}_{\text{b}}\left[U\sum_{i,j}{\!c_{i}c_{j}^{*}\!\left|i\right\rangle \left\langle j\right|\rho_{\text{b}}\left(0\right)}U^{\dagger}\right]\nonumber \\
& \!=\! \sum_{i,j}{\!c_{i}c_{j}^{*}\mathrm{Tr}_{\text{b}}\!\!\left[\rme ^{-\rmi t\left[\!H_{\!+\!}\otimes\left(\!\frac{N_\text{c}}{2}\!+\!S_{z}\!\right)\!+\!H_{\!-\!}\otimes\left(\!\frac{N_\text{c}}{2}\!-\!S_{z}\!\right)\right]}\!\left|i\right\rangle \!\!\left\langle j\right|\!\rho_{\text{b}}\!\left(0\right)\!\rme ^{\rmi t\left[\!H_{\!+\!}\otimes\left(\!\frac{N_\text{c}}{2}\!+\!S_{z}\!\right)\!+\!H_{\!-\!}\otimes\left(\!\frac{N_\text{c}}{2}\!-\!S_{z}\!\right)\right]}\right]}\nonumber \\
& \!=\! \sum_{i,j}{\!c_{i}c_{j}^{*}\mathrm{Tr}_{\text{b}}\!\!\left[\rme ^{-\rmi t\left[H_{\!+\!}\left(\frac{N_\text{c}}{2}\!+\!i\right)\!+\!H_{\!-\!}\left(\frac{N_\text{c}}{2}\!-\!i\right)\right]}\left|i\right\rangle \left\langle j\right|\rho_{\text{b}}\left(0\right)\rme ^{\rmi t\left[H_{\!+\!}\left(\frac{N_\text{c}}{2}\!+\!j\right)\!+\!H_{\!-\!}\left(\frac{N_\text{c}}{2}\!-\!j\right)\right]}\right]}\nonumber \\
& \!=\!  \sum_{i,j}{\!c_{i}c_{j}^{*}\!\left|i\right\rangle \left\langle j\right|\mathrm{Tr}_{\text{b}}\left[\rme ^{-\rmi t\left[i\left(H_{\!+\!}\!-\!H_{\!-\!}\right)\!+\!\frac{N_\text{c}}{2}\left(H_{\!+\!}\!+\!H_{\!-\!}\right)\right]}\rho_{\text{b}}\left(0\right)\rme ^{\rmi t\left[j\left(H_{\!+\!}\!-\!H_{\!-\!}\right)\!+\!\frac{N_\text{c}}{2}\left(H_{\!+\!}\!+\!H_{\!-\!}\right)\right]}\right]}\nonumber \\
&\!=\!  \sum_{i,j}{\!c_{i}c_{j}^{*}\!\left|i\right\rangle \left\langle j\right|\mathrm{Tr}_{\text{b}}\left[\rme ^{-\rmi t\left(2\eta iH_{1}\!+\!H_{0}\left(h\right)\right)}\rho_{\text{b}}\left(0\right)\rme ^{\rmi t\left(2\eta jH_{1}\!+\!H_{0}\left(h\right)\right)}\right]}\nonumber \\
&\! = \! \sum_{i,j}{\!c_{i}c_{j}^{*}\!\left|i\right\rangle \left\langle j\right|\mathrm{Tr}_{\text{b}}\left[\rme ^{\rmi tH_{0}\left(h+2\eta j\right)}\rme ^{-\rmi tH_{0}\left(h+2\eta i\right)}\frac{\rme ^{-\beta H_{0}\left(h\right)}}{Z\left(\beta,h\right)}\right]}\nonumber \\
& \!= \! \sum_{i,j}{\!c_{i}c_{j}^{*}\!\left|i\right\rangle \left\langle j\right|\tilde{\rho}_{ij}\left(t\right)},
\end{align}
\end{widetext}
where the simplified reduced density matrix $\tilde{\rho}_{ij}\!\left(t\right)\!=\!\mathrm{Tr}_{\text{b}}\left[\rme ^{\rmi tH_{0}\left(h+2\eta j\right)}\rme ^{-\rmi tH_{0}\left(h+2\eta i\right)}\right.$ $\left.\times\rme ^{-\beta H_{0}\left(h\right)}/Z\left(\beta,h\right)\right]$.
Here, we obtain the simplified reduced density matrix in the eigenspace $\left\{ \prod_{k>0}{\left|\phi_{l}\right\rangle _{k}};l=0,\pm,2\right\} $ of the original central spin chain $H_{0}\left(h\right)$, where the eigenstates are $\left|\phi_{0}\right\rangle _{k}$, $\left|\phi_{\pm}\right\rangle _{k}=b_{\pm k}^{\dagger}\left|\phi_{0}\right\rangle _{k}$ and $\left|\phi_{2}\right\rangle _{k}=b_{k}^{\dagger}b_{-k}^{\dagger}\left|\phi_{0}\right\rangle _{k}$ and satisfy $b_{\pm k}^{\dagger}b_{\pm k}\left|\phi_{0}\right\rangle _{k}\!=\!b_{\pm k}^{\dagger}b_{\pm k}\left|\phi_{\mp}\right\rangle _{k}\!=\!0$, $b_{\pm k}^{\dagger}b_{\pm k}\left|\phi_{\pm}\right\rangle _{k}\!=\!\left|\phi_{\pm}\right\rangle _{k}$ and $b_{\pm k}^{\dagger}b_{\pm k}\left|\phi_{2}\right\rangle _{k}\!=\!\left|\phi_{2}\right\rangle _{k}$.
It follows that the simplified reduced density matrix is given by
\begin{align}
\!\tilde{\rho}_{ij}\!\left(t\right)  &\!=\!  \mathrm{Tr}_{\text{b}}\left[\rme ^{\rmi tH_{0}\left(h+2\eta j\right)}\rme ^{-\rmi tH_{0}\left(h+2\eta i\right)}\frac{\rme ^{-\beta H_{0}\left(h\right)}}{Z\left(\beta,h\right)}\right]\nonumber \\
&\!=\!  \frac{1}{\! Z\!\left(\!\beta,\!h\!\right)\!}\prod_{k^{\prime}>0}\sum_{l=0,\pm,2}\ _{k^{\prime}}\left\langle \phi_{l}\right|\rme ^{\rmi tH_{0}\left(h+2\eta j\right)}\rme ^{-\rmi tH_{0}\left(h+2\eta i\right)}\nonumber\\
	&\quad\times\rme ^{-\beta H_{0}\left(h\right)}\left|\phi_{l}\right\rangle _{k^{\prime}}\nonumber \\
&\!=\!  \frac{1}{\! Z\!\left(\!\beta,\!h\!\right)\!}\prod_{\!k\!^{\prime}>0}\sum_{l=0,\pm,2}\ _{\!k\!^{\prime}}\!\left\langle \phi_{l}\right|\rme ^{\rmi t\sum_{\!k\!}{E_{j,\!k\!}\left(d_{j,\!k\!}^{\dagger}d_{j,\!k\!}-\!\frac{1}{2}\right)}}\nonumber\\
&\quad\times\rme ^{-\rmi t\sum_{\!k\!}{E_{i,\!k\!}\left(d_{i,\!k\!}^{\dagger}d_{i,\!k\!}-\!\frac{1}{2}\right)}}\rme ^{\!-\!\beta\sum_{\!k\!}{\Lambda_{\!k\!}\left(b_{\!k\!}^{\dagger}b_{\!k\!}-\!\frac{1}{2}\right)}}\!\!\left|\phi_{l}\right\rangle _{\!k\!^{\prime}}\nonumber \\
&\!=\!  \frac{1}{\! Z\!\left(\!\beta,\!h\!\right)\!}\prod_{\! \!k\!>0\!}\sum_{l=0,\pm,2\!}\ _{\!k\!}\!\left\langle \phi_{l}\right|\rme ^{\rmi tE_{j,\!k\!}\left(d_{j,\!k\!}^{\dagger}d_{j,\!k\!}+d_{j,-\!k\!}^{\dagger}d_{j,-\!k\!}-\!1\!\right)}\nonumber\\
&\quad\times\rme ^{\!-\rmi tE_{i,\!k\!}\left(d_{i,\!k\!}^{\dagger}d_{j,\!k\!}+d_{i,-\!k\!}^{\dagger}d_{j,-\!k\!}-\!1\!\right)}\rme ^{\!-\!\beta\Lambda_{\!k\!}\left(b_{\!k\!}^{\dagger}b_{\!k\!}+b_{-\!k\!}^{\dagger}b_{-\!k\!}-\!1\right)}\!\!\left|\phi_{l}\right\rangle _{\!k},
\end{align}
where the effective Hamiltonians $H_{m}\!=\!H_{0}\left(h\!+\!2\eta m\right)$ ($m\!=\!i,j$) can be diagonalized in a similar way as Eq.~(\ref{DiaH}), the angles $\mu_{m,k}\!=\!\arctan\left[\lambda\gamma\sin k/\!\left(\!h\!+\!2\eta m\!-\!\lambda\!\cos k\!\right)\right]$ and $\theta_{m,k}=\left(\mu_{m,k}-\nu_{k}\right)/2$ are determined by the diagonalization conditions, $d_{m,\pm k}\!=\!\cos\!\left(\!\theta_{m,k}\!\right)\!b_{m,\pm\! k}\mp\rmi \sin\!\left(\!\theta_{m,k}\!\right)\!\left(\!b_{m,\mp\! k\!}\right)^{\!\dagger\!}$ and $E_{m,k}\!=\!E_{m,\!-\!k}\!=\!\sqrt{\!\left(\!h\!+\!2\eta m\!-\!\lambda\!\cos\! k\!\right)^{2}\!+\!\lambda^{2}\!\gamma^{2}\!\sin^{2}\!k}$ are the anticommuting fermion operators and the excitation energy, respectively.
By some calculations, we could gain the useful identity relations 
\begin{align}
d_{m,\pm k}^{\dagger}d_{m,\pm k}\left|\phi_{\pm}\right\rangle _{k} & =  \left|\phi_{\pm}\right\rangle _{k},\\
d_{m,\pm k}^{\dagger}d_{m,\pm k}\left|\phi_{\mp}\right\rangle _{k} & =  0,\\
\left(d_{m,\pm k}^{\dagger}d_{m,\pm k}\right)^{n}\left|\phi_{0}\right\rangle _{k} & =  d_{m,\pm k}^{\dagger}d_{m,\pm k}\left|\phi_{0}\right\rangle _{k}\nonumber\\
&=-\rmi \sin\theta_{m,k}\nonumber\\
&\quad\times\left(\cos\theta_{m,k}\left|\phi_{2}\right\rangle _{k}+\rmi \sin\theta_{m,k}\left|\phi_{0}\right\rangle _{k}\right),\\
\left(d_{m,\pm k}^{\dagger}d_{m,\pm k}\right)^{n}\left|\phi_{2}\right\rangle _{k} & = d_{m,\pm k}^{\dagger}d_{m,\pm k}\left|\phi_{2}\right\rangle _{k}\nonumber\\
&=\cos\theta_{m,k}\nonumber\\
&\quad \times\left(\cos\theta_{m,k}\left|\phi_{2}\right\rangle _{k}+\rmi \sin\theta_{m,k}\left|\phi_{0}\right\rangle _{k}\right),
\end{align}
and the expressions 
\begin{align}
\rme ^{-\rmi tE_{m,k}\left(d_{m,k}^{\dagger}d_{m,k}-\frac{1}{2}\right)}\left|\phi_{\pm}\right\rangle _{k} & =  \rme ^{\mp\rmi \frac{tE_{m,k}}{2}}\left|\phi_{\pm}\right\rangle _{k},\label{phi_pm1}\\
\rme ^{-\rmi tE_{m,k}\left(d_{m,-k}^{\dagger}d_{m,-k}-\frac{1}{2}\right)}\left|\phi_{\pm}\right\rangle _{k} & = \rme ^{\pm\rmi \frac{tE_{m,k}}{2}}\left|\phi_{\pm}\right\rangle _{k},\label{phi_pm2}
\end{align}
\begin{align}
&\quad \rme ^{\pm\rmi tE_{m,k}\left(d_{m,k}^{\dagger}d_{m,k}-\frac{1}{2}\right)}\left|\phi_{0}\right\rangle _{k} \nonumber\\
& =  \left(\rme ^{\mp\rmi \frac{tE_{m,k}}{2}}\pm2\rmi \sin^{2}\theta_{m,k}\sin\frac{E_{m,k}t}{2}\right)\left|\phi_{0}\right\rangle _{k}\nonumber\\
&\quad\pm\sin2\theta_{m,k}\sin\frac{E_{m,k}t}{2}\left|\phi_{2}\right\rangle _{k},\label{phi_0}\\
&\quad \rme ^{\pm\rmi tE_{m,k}\left(d_{m,k}^{\dagger}d_{m,k}-\frac{1}{2}\right)}\left|\phi_{2}\right\rangle _{k} \nonumber\\
& =  \mp\sin2\theta_{m,k}\sin\frac{E_{m,k}t}{2}\left|\phi_{0}\right\rangle _{k}\nonumber\\
&\quad+\left(\rme ^{\mp\rmi \frac{tE_{m,k}}{2}}\pm2\rmi \cos^{2}\theta_{m,k}\sin\frac{E_{m,k}t}{2}\right)\left|\phi_{2}\right\rangle _{k}.\label{phi_2}
\end{align}
From Eqs.~(\ref{phi_pm1}) to (\ref{phi_2}), we obtain the inner product results 
\begin{widetext}
\begin{align}
& \   \quad_{k}\left\langle \phi_{\pm}\right|\rme ^{\rmi tE_{j,k}\left(d_{j,k}^{\dagger}d_{k}+d_{j,-k}^{\dagger}d_{-k}-1\right)}\rme ^{-\rmi tE_{i,k}\left(d_{i,k}^{\dagger}d_{k}+d_{i,-k}^{\dagger}d_{-k}-1\right)}\rme ^{-\beta\Lambda_{k}\left(b_{k}^{\dagger}b_{k}+b_{-k}^{\dagger}b_{-k}-1\right)}\left|\phi_{\pm}\right\rangle _{k}\nonumber \\
& =  \,_{k}\left\langle \phi_{\pm}\right|\rme ^{\rmi tE_{j,k}\left(1-1\right)}\rme ^{-\rmi tE_{i,k}\left(1-1\right)}\rme ^{-\beta\Lambda_{k}\left(1-1\right)}\left|\phi_{\pm}\right\rangle _{k}\nonumber \\
& =  1.\label{InnerProdpm}\\
& \ \quad  _{k}\left\langle \phi_{0}\right|\rme ^{\rmi tE_{j,k}\left(d_{j,k}^{\dagger}d_{k}+d_{j,-k}^{\dagger}d_{-k}-1\right)}\rme ^{-\rmi tE_{i,k}\left(d_{i,k}^{\dagger}d_{k}+d_{i,-k}^{\dagger}d_{-k}-1\right)}\rme ^{-\beta\Lambda_{k}\left(b_{k}^{\dagger}b_{k}+b_{-k}^{\dagger}b_{-k}-1\right)}\left|\phi_{0}\right\rangle _{k}\nonumber \\
& =  \rme ^{\beta\Lambda_{k}}\,_{k}\left\langle \phi_{0}\right|\rme ^{\rmi tE_{j,k}\left(d_{j,k}^{\dagger}d_{k}+d_{j,-k}^{\dagger}d_{-k}-1\right)}\rme ^{-\rmi tE_{i,k}\left(d_{i,k}^{\dagger}d_{k}+d_{i,-k}^{\dagger}d_{-k}-1\right)}\left|\phi_{0}\right\rangle _{k}\nonumber \\
& =  \rme ^{\beta\Lambda_{k}}\,_{k}\left\langle \phi_{0}\right|\rme ^{2\rmi tE_{j,k}\left(d_{j,k}^{\dagger}d_{k}-\frac{1}{2}\right)}\rme ^{-2\rmi tE_{i,k}\left(d_{i,k}^{\dagger}d_{k}-\frac{1}{2}\right)}\left|\phi_{0}\right\rangle _{k}\nonumber \\
& =  \rme ^{\beta\Lambda_{k}}\left\{ _{k}\left\langle \phi_{0}\right|\left[\rme ^{-\rmi tE_{j,k}}+2\rmi \sin^{2}\theta_{j,k}\sin\left(tE_{j,k}\right)\right]-_{k}\left\langle \phi_{2}\right|\left[\sin2\theta_{j,k}\sin\left(tE_{j,k}\right)\right]\right\} \nonumber \\
& \quad  \times\left\{ \left[\rme ^{\rmi tE_{i,k}}-2\rmi \sin^{2}\theta_{i,k}\sin\left(tE_{i,k}\right)\right]\left|\phi_{0}\right\rangle _{k}-\sin2\theta_{i,k}\sin\left(tE_{i,k}\right)\left|\phi_{2}\right\rangle _{k}\right\} \nonumber \\
& =  \rme ^{\!\beta\Lambda_{k}}\!\left\{ \left[\rme ^{-\rmi tE_{j,k}\!}\!+\!2\rmi \sin^{2}\!\theta_{j,k}\sin\left(tE_{j,k}\!\right)\right]\left[\rme ^{\rmi tE_{i,k}\!}\!-\!2\rmi \sin^{2}\!\theta_{i,k}\sin\left(tE_{i,k}\!\right)\right]\!+\!\sin2\theta_{i,k}\sin2\theta_{j,k}\sin\left(tE_{i,k}\!\right)\sin\left(tE_{j,k}\!\right)\right\} \!,\label{InnerProd0}\\
& \ \quad  _{k}\left\langle \phi_{2}\right|\rme ^{\rmi tE_{j,k}\left(d_{j,k}^{\dagger}d_{k}+d_{j,-k}^{\dagger}d_{-k}-1\right)}\rme ^{-\rmi tE_{i,k}\left(d_{i,k}^{\dagger}d_{k}+d_{i,-k}^{\dagger}d_{-k}-1\right)}\rme ^{-\beta\Lambda_{k}\left(b_{k}^{\dagger}b_{k}+b_{-k}^{\dagger}b_{-k}-1\right)}\left|\phi_{2}\right\rangle _{k}\nonumber \\
& =  \rme ^{-\beta\Lambda_{k}}\,_{k}\left\langle \phi_{2}\right|\rme ^{\rmi tE_{j,k}\left(d_{j,k}^{\dagger}d_{k}+d_{j,-k}^{\dagger}d_{-k}-1\right)}\rme ^{-\rmi tE_{i,k}\left(d_{i,k}^{\dagger}d_{k}+d_{i,-k}^{\dagger}d_{-k}-1\right)}\left|\phi_{2}\right\rangle _{k}\nonumber \\
& =  \rme ^{-\beta\Lambda_{k}}\,_{k}\left\langle \phi_{2}\right|\rme ^{2\rmi tE_{j,k}\left(d_{j,k}^{\dagger}d_{k}-\frac{1}{2}\right)}\rme ^{-2\rmi tE_{i,k}\left(d_{i,k}^{\dagger}d_{k}-\frac{1}{2}\right)}\left|\phi_{2}\right\rangle _{k}\nonumber \\
& =  \rme ^{-\beta\Lambda_{k}}\left\{ _{k}\left\langle \phi_{0}\right|\left[\sin2\theta_{j,k}\sin\left(tE_{j,k}\right)\right]+_{k}\left\langle \phi_{2}\right|\left[\rme ^{-\rmi tE_{j,k}}+2\rmi \cos^{2}\theta_{j,k}\sin\left(tE_{j,k}\right)\right]\right\} \nonumber \\
& \quad  \times\left\{ \sin2\theta_{i,k}\sin\left(tE_{i,k}\right)\left|\phi_{0}\right\rangle _{k}+\left[\rme ^{\rmi tE_{i,k}}-2\rmi \cos^{2}\theta_{i,k}\sin\left(tE_{i,k}\right)\right]\left|\phi_{2}\right\rangle _{k}\right\} \nonumber \\
& =  \rme ^{\!-\!\beta\Lambda_{k}\!}\!\left\{ \left[\rme ^{-\rmi tE_{j,k}\!}\!+\!2\rmi \cos^{2}\!\theta_{j,k}\sin\left(tE_{j,k}\!\right)\!\right]\!\left[\rme ^{\rmi tE_{i,k}\!}\!-\!2\rmi \cos^{2}\!\theta_{i,k}\sin\left(tE_{i,k}\!\right)\!\right]\!+\!\sin2\theta_{i,k}\sin2\theta_{j,k}\sin\left(tE_{i,k}\!\right)\sin\left(tE_{j,k}\!\right)\right\} \!,\label{InnerProd2}
\end{align}
\end{widetext}

From Eqs.~(\ref{InnerProd0}) to (\ref{InnerProd2}), we could gain the reduced density matrix of the central spins (presented as Eq.~(\ref{Final StateContext}) in the main text)
\begin{widetext} 
\begin{align}
\rho_{\text{c}}\left(t\right) & =  \sum_{i,j}{c_{i}c_{j}^{*}\left|i\right\rangle \left\langle j\right|\tilde{\rho}_{ij}\left(t\right)}\nonumber \\
& =  \sum_{i,j}\bigg\{\frac{c_{i}c_{j}^{*}}{Z\left(\beta,h\right)}\left|i\right\rangle \left\langle j\right|\prod_{k>0}\left[2+2\cosh\left(\beta\Lambda_{k}\right)C\left(\theta_{i,k},E_{i,k}t\right)C\left(\theta_{j,k},E_{j,k}t\right)\right.\nonumber \\
& \quad  \left.+\rme ^{\beta\Lambda_{k}}A\left(\theta_{j,k},E_{j,k}t\right)A^{*}\left(\theta_{i,k},E_{i,k}t\right)+\rme ^{-\beta\Lambda_{k}}B\left(\theta_{j,k},E_{j,k}t\right)B^{*}\left(\theta_{i,k},E_{i,k}t\right)\right]\bigg\},\label{Final State}
\end{align}
\end{widetext}
where the coefficients are defined as $A\left(X,Y\right)=\rme ^{-\rmi Y}+2\rmi \sin^{2}X\sin Y$, $B\left(X,Y\right)=\rme ^{-\rmi Y}+2\rmi \cos^{2}X\sin Y$, $C\left(X,Y\right)=\sin2X\sin Y$ and $X$,$Y$ are real variables.
With the reduced density matrix, one could calculate the time evolution of numerous intriguing quantities about the central spins.

\end{appendix}



%

\end{document}